\documentclass{ar-1col}
\usepackage{natbib}
\usepackage[mathscr]{euscript}
\usepackage{mathtools}
\usepackage{upgreek}

\usepackage{color}
\usepackage{float}

\jname{Xxxx. Xxx. Xxx. Xxx.}
\jvol{AA}
\jyear{2022}
\doi{10.1146/((please add article doi))}

\newcommand{\simkl}{\stackrel{{<}}{{\scriptstyle\sim}}}

\begin{document}

\markboth{Kurtz}{Asteroseismology across the HR diagram}

\title{Asteroseismology across the HR diagram}

\author{Donald Kurtz$^{1,2}$ 
\affil{$^{1}$Department of Physics, North-West University, Dr Albert Luthuli Drive, Mahikeng 2735, South Africa}
\affil{$^{2}$Jeremiah Horrocks Institute, University of Central Lancashire, Preston PR1~2HE, UK}
\affil{ email: kurtzdw@gmail.com; ORCID: 0000-0002-1015-3268}
}

\begin{abstract}
Asteroseismology has grown from its beginnings three decades ago to a mature field teeming with discoveries and applications. This phenomenal growth has been enabled by space photometry with precision $10-100$ times better than ground-based observations, with nearly continuous light curves for durations of weeks to years, and by large scale ground-based surveys spanning years designed to detect all time-variable phenomena.   The new high precision data are full of surprises, deepening our understanding of the physics of stars. 

\hangindent=.3cm$\bullet$ This review explores asteroseismic  developments from the last decade primarily as a result of light curves from the {\it Kepler} and TESS space missions for: massive upper main-sequence OBAF stars, pre-main-sequence stars, peculiar stars, classical pulsators, white dwarfs and subdwarfs, and tidally interacting close binaries.

\hangindent=.3cm$\bullet$ The space missions have increased the numbers of pulsators in many classes by an order of magnitude.

\hangindent=.3cm$\bullet$ Asteroseismology measures fundamental stellar parameters and stellar interior physics -- mass, radius, age, metallicity, luminosity, distance, magnetic fields, interior rotation, angular momentum transfer, convective overshoot, core burning stage --  supporting disparate fields such as galactic archeology, exoplanet host stars, supernovae progenitors, gamma ray and gravitational wave precursors, close binary star origins and evolution, and standard candles.

\hangindent=.3cm$\bullet$ Stars are the luminous tracers of the universe. Asteroseismology significantly improves models of stellar structure and evolution on which all inference from stars depends. 
\end{abstract}

\begin{keywords}
asteroseismology,  
stars: oscillations (including pulsations),
stars: interiors,
stars: variables: general,
stars: binaries: close
\end{keywords}
\maketitle

\tableofcontents

\section{Introduction}

When I was an undergraduate at San Diego State University studying astronomy in the late 1960s, I had then on my bookshelf the classic work on Stellar Structure, {\it The Internal Constitution of the Stars} \citep{1926ics..book.....E}, which I  had purchased used for a mere 50 cents. It is on my bookshelf to this day. \citet{1920Obs....43..341E} is credited with being the first to propose hydrogen fusion as the power source for stars. He developed that idea further in {\it The Internal Constitution of the Stars}, although he was reluctant to admit a hydrogen fraction greater than 7\%, which made his conclusions marginal as to whether there was enough hydrogen in the Sun to power it for the time required by Earth's geological record. His reluctance was despite Cecilia \citet{1925PhDT.........1P} having shown in her work at Harvard College Observatory that stars are composed primarily of hydrogen and helium, as presented in her Radcliffe College PhD thesis. She was English, but studied at Harvard, since Cambridge, where Eddington was,  did not award PhDs to women at that time. \citet{1962asce.book.....S} declared her thesis to be ``undoubtedly the most brilliant PhD thesis ever written in astronomy.'' Perhaps Eddington was influenced by Henry Norris Russell (the R in HR diagram), who believed then that the stars had compositions similar to those of the Earth. He did not concede that Payne was correct until 1929.

In  {\it The Internal Constitution of the Stars} Eddington began with these words: 
\begin{extract}
\noindent At first sight it would seem that the deep interior of the Sun and stars is less accessible to scientific investigation than any other region of the universe. Our telescopes may probe farther and farther into the depths of space; but how can we ever obtain certain knowledge of that which is hidden behind substantial barriers? What appliance can pierce through the outer layers of a star and test the conditions within?
\end{extract}
\noindent Eddington's answer to this was theory. Nevertheless, in his third paragraph of that opening page, he warned: ``We should be unwise to trust scientific inference very far when it becomes divorced from opportunity for observational test.'' This interesting contention both binds and divides astrophysicists across the breadth of our all-encompassing field. 

It is more than 50 years since I first read Eddington's treatise, at a time when it was then only 40 years since he had written about such uncertainties in our understandings of the physics of the interior of stars. As I finished my undergraduate studies and moved to the University of Texas in 1970 for my PhD, I heard from my fellow students about the ``solar 5-minute oscillations". These were first noted in a broad report on the  velocity fields in the solar atmosphere by \citet{1962ApJ...135..474L}. Then \citet{1970ApJ...162..993U} and \citet{1971ApL.....7..191L} correctly identified the 5-minute oscillations as standing acoustic waves. The breakthrough came just a few years later when \citet{1975A&A....44..371D} resolved ``three or four discrete stable modes'' in a plot of wavenumber versus frequency. Many other studies followed and Helioseismology was born. (See \citet{2021LRSP...18....2C} for a comprehensive history and account of solar structure and evolution, including helioseismology.) Eddington's appliance that could pierce through the outer layers of a star had been found, at least for the Sun. 

Astronomers are often asked: What good is astronomy? That question comes particularly from funding agencies wanting to know what economic benefit or other ``impact'' there might be. This question reminds me of Howard Florey, who shared the 1945 Nobel Prize in Medicine for bringing penicillin into medical use, saving untold millions of lives. Talk about impact! Yet when asked about his work Florey famously said, 

\begin{extract}
\noindent People sometimes think that I and the others worked on penicillin because we were interested in suffering humanity. I don't think it ever crossed our minds about suffering humanity. This was an interesting scientific exercise, and because it was of some use in medicine is very gratifying, but this was not the reason that we started working on it\footnote{De Berg Collection in the National Library of Australia.}.   
\end{extract} 

Three decades after helioseismology, the announcement of the birth of asteroseismology was made  by \citet{2001Sci} as a result of the discovery of solar-like oscillations in the G2V solar analogue $\alpha$\,Cen\,A by \citet{2001A&A...374L...5B} and in the G2IV subgiant star $\beta$\,Hyi by \citet{2001ApJ...549L.105B}. In the more than two decades since then, asteroseismology has blossomed, with major impact on our understanding of stellar pulsation, stellar structure and evolution, and with wide applications in other fields of astrophysics. Nevertheless, in the spirit of Howard Florey, asteroseismologists are largely driven by a passion to understand the stars, although they are also gratified by applications of their results to other fields.

\subsection{The scope of this review}

The greatest success of asteroseismology is, of course, helioseismology. With spatially resolved observations, the interior structure of the Sun is constrained by measurements of the square of the sound speed to better than 3 parts per thousand from surface to core, and much better than that throughout most of the Sun. Nothing in asteroseismology yet competes with that. Nevertheless, there have been great successes in asteroseismic studies of stochastically-driven pulsators -- the solar-like and red giant oscillators -- for which pulsations are driven by broad-spectrum white noise generated by convection.  These studies have led to high accuracy (less than a few percent) determinations of stellar mass and radius, knowledge that is critical to exoplanet studies. They have provided direct views -- in the Eddington sense -- of interior rotation and the extent of the convective core boundary, with impact on stellar structure and evolution models. They have made it possible to distinguish -- all jumbled together in the HR~Diagram -- which red giants are core helium burning and which are hydrogen shell-burning. They have given model-dependent age determinations for single stars (cluster stellar age determinations are also model-dependent) that are useful for galactic archeology \citep{2017AN....338..644M}.  

These successes are not discussed further here, as the solar-like oscillators have been thoroughly reviewed recently by \citet{2013ARA&A..51..353C}, \citet{2019LRSP...16....4G}, and \citet{2021FrASS...7..102J}. \citet{2017A&ARv..25....1H} and \citet{2020FrASS...7...44B} have reviewed asteroseismic red giant discoveries, while \citet{2021ApJ...919..131H} have found over 158\,000 pulsating red giants, an order of magnitude increase on the already huge number found by {\it Kepler} and K2.  \citet{2021RvMP...93a5001A} and \citet{2019ARA&A..57...35A} examined results for interior rotation in over 1200 giants  and main-sequence stars\footnote{A number that has increased by nearly 700 (\citealt[][for $\gamma$~Dor stars]{2020MNRAS.491.3586L}; \citealt[][for core He-burning red giants]{2019ApJ...887..203T}), showing how rapidly this field is expanding.}, showing that these are consistent with white dwarf rotation rates. Interestingly, \citet{2020MNRAS.499.4687C} provide a unifying description of the transition from the stochastically driven solar-like and red giant oscillators through stars with both stochastically driven and classical heat-engine driven pulsations among the semi-regular variables, to the classical pulsations of the Mira variables. 

This review looks in the era of space photometry at the observational discoveries and advances in asteroseismology for pulsating stars other than the solar-like and red giant stars. This means OBAF stars of the upper main-sequence, subdwarfs, white dwarfs, and giant stars in the classical instability strip. The discussion here is based primarily on the vast, spectacularly successful data sets of the {\it Kepler} and TESS missions. They were not the first, or only, asteroseismic satellites. Other missions have had significant successes. Both \citet{2020FrASS...7...70B} and \citet{2021RvMP...93a5001A} discuss the asteroseismic space missions earlier than {\it Kepler}: WIRE, MOST, CoRoT, as well as the contemporary nano-satellite cluster BRITE. 

In the appendix I provide some personal stories and historical background, along with extensive detailed sections for various classes of pulsating stars studied asteroseismically.

\subsection{Asteroseismology: terminology and a pulsation HR~Diagram}

The basic data of asteroseismology are pulsation mode frequencies. Model frequencies are tuned by varying the internal physics to match observed frequencies in a wide spectrum of models with different input parameters, constraining the interior structure of the star. This is referred to as forward modeling, to distinguish it from inversion of the mode frequencies as is done for the Sun. Open-source stellar structure and evolution models from MESA (Modules for Experiments in Stellar Astrophysics; e.g., \citealt{2019ApJS..243...10P}) in conjunction with the pulsation code GYRE \citep{2013MNRAS.435.3406T} are widely used in asteroseismology to constrain models of stellar structure, hence also evolution.  The theory of asteroseismology has been reviewed in depth by  \citet{2021RvMP...93a5001A} and  \citet[chapter 3]{2010aste.book.....A}. I provide here only the terminology necessary for this observational review. 

\begin{marginnote}[]
\entry{Forward modeling} \  The asteroseismic method of matching model pulsation mode frequencies with observed frequencies with known mode identification, $n,\ell,m$, to constrain stellar interior physics. 
\end{marginnote}

The heat engine pulsators are stars for which pulsation driving is stronger than damping and for which the pulsation time is much less than the thermal time of the star. Driving is mostly by an opacity mechanism,   the $\kappa$-mechanism, operating in an ionization zone of an abundant ion -- usually H or He, but also, e.g.,  Fe-Ni in upper main-sequence stars, and C-O in white dwarfs. The $\kappa$-mechanism can be accompanied by other mechanisms: variations in the adiabatic exponent, $\gamma$, tidal excitation, and, potentially, variation in the nuclear energy generation rate, the $\epsilon$-mechanism. The heat engine pulsators generally have mode lifetimes that are long compared to our observation times. Stars for which damping exceeds such driving can still be driven to observable pulsation by stochastic driving  -- as in the solar-like stars (giants and red giants that are not a topic in this review) -- but also in some AF stars and all massive OB stars, a new asteroseismic discovery (Section\,\ref{slf}). 

There are five restoring forces: pressure (p\,modes), buoyancy (gravity, or g\,modes), Coriolis force (inertial and Rossby, r~modes), magnetic (Alfv\'en), and tidal.  These forces can, and often do, operate together, but the main restoring forces are usually pressure and buoyancy. Coupled g\,modes and p\,modes are referred to as mixed modes. When the Coriolis force acts in concert with gravity, the modes are called gravito-inertial modes. When there is a strong, global magnetic field, its resulting Lorentz force along with the pressure force gives rise to magneto-acoustic modes. With the superb precision of the space photometry, it is now possible to observe the effects of these many restoring forces and to extract new asteroseismic inference from those effects. 

To proceed with forward modeling it is necessary to identify the modes of all frequencies modeled. Those are described in terms of spherical harmonics, even though those functions are only truly appropriate for perfectly spherical stars that are not rotating and have no magnetic fields. Nevertheless, the terminology is useful. See \citet{2021RvMP...93a5001A} and \citet[chapter 1]{2010aste.book.....A} for further introduction to this. For this review the three quantum numbers describing the spherical harmonics are the radial overtone, $n$; the degree, $\ell$, and the azimuthal order, $m$. Radial modes have $\ell = 0$, nonradial modes have $\ell > 0$. The principal nonradial modes observed are dipole ($\ell = 1$) and quadrupole ($\ell = 2$) modes, since cancellation makes $\ell \ge 3$ modes undetectable in most cases. Modes with $m = 0$ are zonal modes; with $m = | \ell |$ are sectoral modes; with $0 < m < | \ell |$ are tesseral modes. The sectoral modes primarily, but potentially also the tesseral modes, are of great interest, since these are traveling waves with observed frequencies that depend on rotation in the mode cavities, hence they provide the data on interior rotation, information that is unobtainable by any method other than asteroseismology.  

\begin{marginnote}[]
\entry{Spherical harmonic quantum numbers, $n, \ell, m$} \  $n =$ radial overtone; $\ell =$ degree; \\\ $m =$ azimuthal order 
\entry{The large frequency separation} \  $\Delta \nu$: the inverse of the sound travel time across the star; gives a measure of the mean density.  
\entry{The characteristic period} \  $\Pi_0$: the buoyancy travel time across the star.
\end{marginnote}

Note that:  p\,modes have largely radial displacement, g\,modes horizontal; p\,modes are primarily sensitive to the conditions in the outer envelope of the star, g\,modes to the deeper interior; p\,modes have higher frequencies, g\,modes lower frequencies;  p\,mode frequencies increase with increasing radial overtone $n$, g\,modes frequencies decrease with increasing $n$; high overtone p\,mode frequencies asymptotically approach uniform spacing dependent on the large separation, $\Delta \nu$; high overtone g\,mode periods asymptotically approach uniform spacing dependent on the characteristic period, $\Pi_0$.

\begin{figure}[t]
\includegraphics[width=1.0\textwidth]{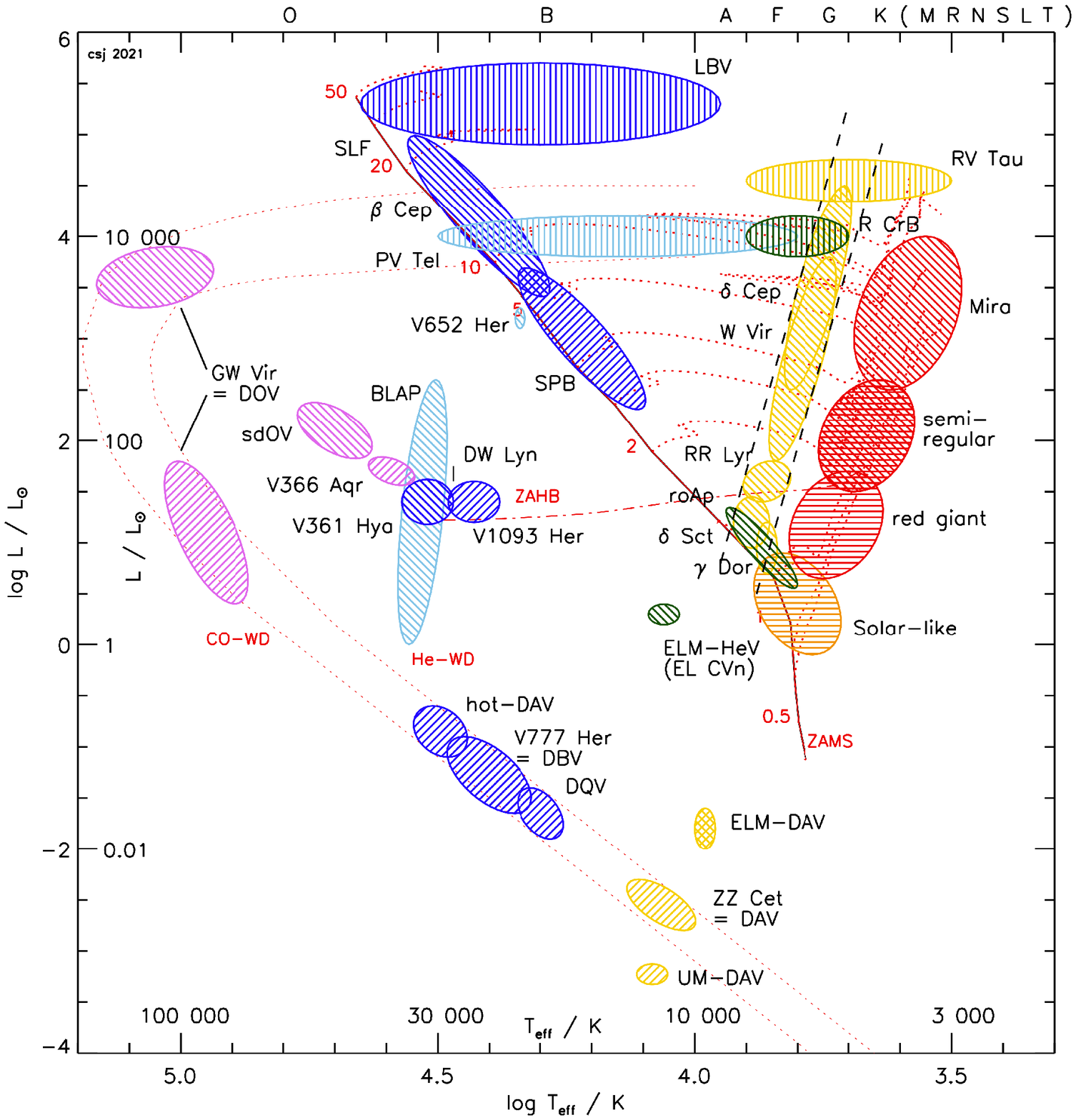}
\caption{A pulsation, or asteroseismic, HR~Diagram. This gives a schematic guide to classes of pulsating stars. The abbreviations for the names of many of the classes follow that of \citet[chapter 2]{2010aste.book.....A}. Others are clarified in the relevant sections of the text. The solid red line is the zero-age-main-sequence, and standard evolutionary tracks are shown as dotted lines up to the tip of the first red giant stage; the red numbers along the main-sequence are masses in M$_\odot$. The cross-hatchings represent the primary mode types: acoustic p\,modes ($\backslash\backslash\backslash$); gravity (buoyancy) g\,modes ($///$); stochastically driven pulsators ($\equiv$); and strange modes ($|||$). This figure is courtesy of Simon Jeffery, and is based on Figure\,1 of \citet[][``Subaru and Swift observations of V652\,Herculis: resolving the photospheric pulsation'']{2015MNRAS.447.2836J}. }
\label{fig:hrd}
\end{figure} 

The pulsation HR~Diagram in Figure\,\ref{fig:hrd} gives an overview of the classes of pulsating stars. This diagram was initiated by \citet{1998Ap&SS.261....1C} and has been adapted many times. Versions can be found in  \citet{2010aste.book.....A}, \citet{2016MNRAS.458.1352J},  \citet[]{2016A&G....57d4.37K}, \citet{2010PhDT.......226D}, \citet{2013PhDT.........6P}, and \citet{2021RvMP...93a5001A}. This diagram is a useful guide to our discussion, but is intentionally schematic, since the classes of the stars and their boundaries are fluid. Taxonomy is useful for organizing our thoughts, but species cross-breed, have subspecies, appear in the record and go extinct. The sections of this review clarify the details. 

\subsection{Frequency analysis}
\label{freqanalysis}

The observations made for asteroseismology are time series of stellar variability, either photometric variations in brightness, spectroscopic radial velocities, or line profile variations. The main observational goal is to extract the mode frequencies and identify the modes ($n,\ell,m$) to then match with model mode frequencies. The emphasis in this review is on the space revolution, from which the frequencies are extracted from light curves by means of Fourier transforms, giving power spectra, power density spectra, amplitude spectra, frequency spectra, or periodograms. For the heat engine pulsators the amplitude spectrum is most widely used, but for succinctness I use here the generic term FT (Fourier transform). See \citet[chapter 5]{2010aste.book.....A} for an in-depth discussion. 

\begin{marginnote}[]
\entry{FT = Fourier transform; Discrete Fourier transform} \  FT is used generically to mean a power spectrum, power density spectrum, amplitude spectrum, frequency spectrum, or periodogram.
\end{marginnote}

\begin{figure*}[t]
\centering
\includegraphics[width=1.0\linewidth,angle=0]{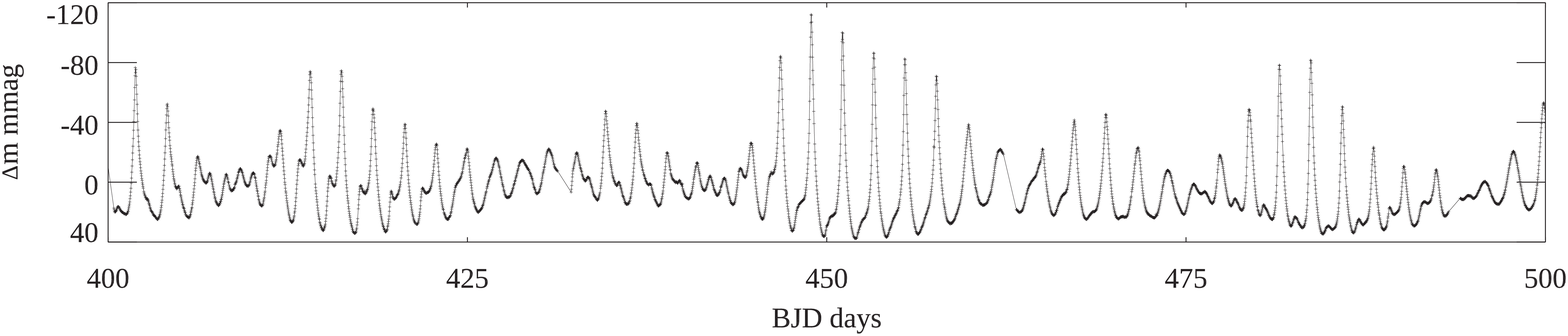}
\includegraphics[width=1.0\linewidth,angle=0]{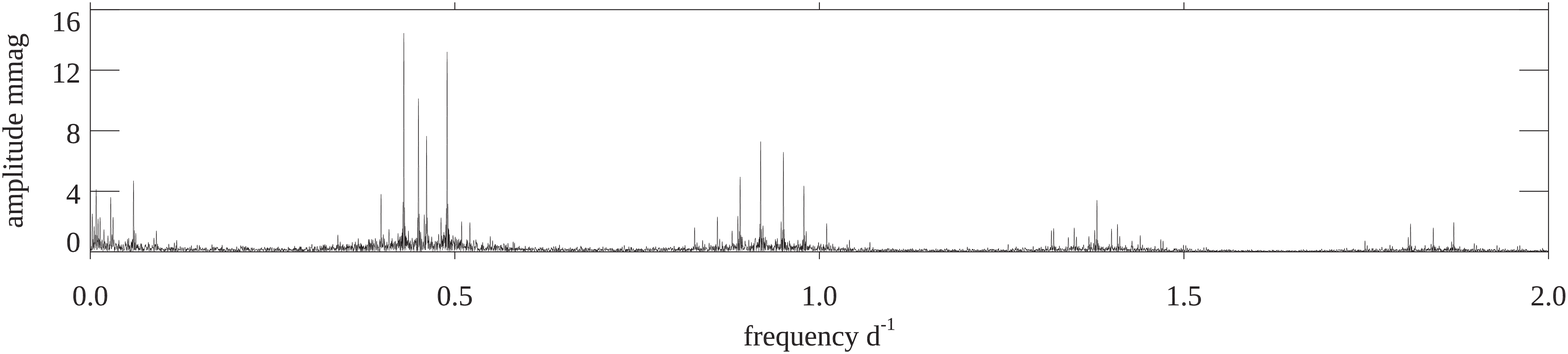}
\caption{
The top panel shows a light curve for KIC\,8113452, a strongly nonlinearly pulsating $\gamma$~Dor star. The time is relative to BJD\,2455000. This light curve is shown here as a clear example of the time series that are the fundamental observational data of asteroseismology. The {\it Kepler} and TESS data are so precise that for many kinds of multi-periodic pulsating stars the light curve can seem a jumble of points, even though at the scale of this plot the uncertainty in magnitude is 1000 times smaller than the size of the data points. The bottom panel shows an FT of this light curve in the low-frequency range of its g~modes. The groups of frequencies around $0.4 -0.5$\,d$^{-1}$ are from the g~modes. The other groups of frequencies are nonlinear combination frequencies that result from the coupled nonlinear pulsations.  It is the job of the asteroseismologist to identify the correct mode frequencies for forward modeling from the plethora of combination frequencies. This is often done by pattern recognition, which humans are good at. But the big data of modern asteroseismology from space missions is overwhelming humans, and machine learning now usually makes the initial classification of the type of pulsator and flags objects of interest. The bottom panel is adapted from Figure 3 of \citet[][``A unifying explanation of complex frequency spectra of {\ensuremath{\gamma}} Dor, SPB and Be stars: combination frequencies and highly non-sinusoidal light curves'']{2015MNRAS.450.3015K}.}
\label{fig:lc-ft}
\end{figure*}

Figure\,\ref{fig:lc-ft} shows examples of a light curve and its FT. Almost all primary publications in observational asteroseismology present FTs of the data under discussion. \citet[chapter 2]{2010aste.book.....A} show examples of ground-based light curves and FTs of most of the classes discussed in this review, and I refer the reader to the referenced primary literature and abundant recent specialist reviews for the space-mission light curves and FTs for each class of pulsating stars. 

\subsection{Photometric precision and duty cycle}

To extract mode frequencies from FTs, we seek high signal-to-noise, high frequency resolution, long time span, and continuous data. Under ground-based photometric conditions, noise is from instrumental variations, atmospheric transparency variations, photon statistics, and scintillation, the latter two of which are reduced with larger aperture telescopes. Space photometry eliminates transparency and scintillation noise. We seek longer time spans of observations for sharper frequency resolution, and we seek better duty cycles (the fraction of time on target) to minimize confusion from complex, overlapping spectral windows in the FT. For a thorough discussion of observational techniques and frequency analysis see \citet[chapters 4 and 5]{2010aste.book.....A}. 

In the 1980s I was making precise ground-based photometric measurements of stellar brightness, reaching photometric precision in pulsation amplitude of $10^{-4}$ in intensity, or 100\,$\upmu$mag. But I could only do that: 1) when observing at night, so the time series had daily gaps, and the FTs had daily aliases;  2) when the telescope was scheduled for me, leading to monthly aliases, since photometry of bright stars was scheduled for bright moon; 3) when the star was not too close to the Sun, leading to yearly aliases in long data sets, and 4) when the weather was good! Long-term weather averages for the superb photometric conditions needed for asteroseismology are typically about $50 - 70$\% for excellent observing sites. 

In one case, the Whole Earth Telescope (WET), with 46 astronomers working over 35\,d, obtained a 35\% duty cycle and reached a photometric precision of 14\,$\upmu$mag \citep{2005MNRAS.358..651K} -- but for only one star! Compare that to the {\it Kepler} main mission where $\sim$150\,000 stars were observed nearly continuously for 4\,yr, and $\sim$50\,000 more were observed for at least 90\,d, with a best precision of the order of 1\,$\upmu$mag -- 1\,ppm. Compare it to the all-sky photometry of TESS with at least 27-d light curves at a precision of order 10\,$\upmu$mag. With an improvement in precision typically of a factor of $10 -100$ over excellent ground-based data, and with data sets that are nearly continuous for time spans of months to years, vast numbers of light curves hold discoveries. Tycho Brahe's quest for ever higher astrometric precision was fundamental to Johannes Kepler's discovery of his laws of motion for the planets (which led to, much later, the naming of the {\it Kepler} mission). \citet{2016A&G....57d4.37K},  in the spirit of Tycho Brahe, called this quest for precision the Tychonic Principle, where a revolutionary improvement in precision inevitably leads to discovery. 

For stars pulsating in g\,modes and r~modes, time spans of at least months are needed to resolve the pulsation frequencies. That resolution is critical to mode identification and modeling of the stellar interior. For the stars that need these long time spans, the 4-yr {\it Kepler} data are currently definitive. For stars for which shorter time spans are sufficient, the TESS mission data are the foundation of much current research. These space missions have revolutionized asteroseismology.

\subsection*{Summary: The space revolution}

\begin{enumerate}
\item Space photometry has precision about 100 times better than ground-based photometry in the range of pulsation frequencies of most stars studied asteroseismically.
\item The space data are nearly continuous, greatly reducing -- or even eliminating --  confusion from spectral window aliases in the FT caused by gaps in the data.
\item The space data -- particularly that of {\it Kepler}, and of TESS in the continuous viewing zone -- have the high frequency resolution that is needed for mode identification and to disentangle the FTs of many types of pulsating stars. 
\item The space data have overwhelmed previous ground-based asteroseismology. For example, the number of $\delta$~Sct and $\gamma$~Dor stars known prior to the space data was a few hundred. It is now many thousands. For solar-like and red giant pulsators it was dozens, now the number surpasses 100\,000. For stars with known internal rotation it was one -- the Sun. It is now thousands. 
\item As a consequence of the discoveries from the space data, there has been a flood of recent reviews, showing the vibrancy of this field of research: 
\end{enumerate}
\begin{itemize} 
 \item This review complements the theoretical asteroseismology review of \citet{2021RvMP...93a5001A} and updates  \citet[chapter 2]{2010aste.book.....A}.
\item \citet{2019ARA&A..57...35A} reviewed angular momentum transport in stellar interiors.
\item \citet{2013ARA&A..51..353C}, \citet{2019LRSP...16....4G}, and \citet{2021FrASS...7..102J} reviewed the stochastic solar-like oscillators. 
\item \citet{2021LRSP...18....2C} reviewed solar structure, evolution, and helioseismology. 
\item \citet{2017A&ARv..25....1H} and \citet{2020FrASS...7...44B} reviewed asteroseismology of red giants.
\item  \citet{2020FrASS...7...70B, 2020svos.conf...53B} reviewed OB stars and their pulsations. 
\item \citet{2021FrASS...8...55G} reviewed {\it Kepler} and TESS observations of $\gamma$~Dor and $\delta$~Sct stars.
\item \citet{2021FrASS...8...31H} reviewed {\it Kepler} observations of roAp stars. 
\item \citet{2021FrASS...7...81P} reviewed {\it Kepler} observations of RR\,Lyrae stars.
\item \citet{2019A&ARv..27....7C} reviewed pulsating white dwarfs, in general, and \citet{2020FrASS...7...47C} reviewed white dwarf asteroseismology with the Kepler telescope.
\item \citet{2016PASP..128h2001H} reviewed sdB and sdO stars.
\item \citet{2021FrASS...8...19L} reviewed asteroseismology of hot subdwarf stars. 
\item \citet{2021FrASS...8...67G} reviewed asteroseismology of close binary stars, and \citet{2021Galax...9...28L} reviewed eclipsing binaries with pulsating components.
\end{itemize}

\section{Pulsation in main-sequence OBAF stars: $\gamma$~Dor, $\delta$~Sct, roAp, SPB, and $\beta$\,Cep stars}
\label{obaf}

Along the main-sequence in the AF range ($M = 1.3 - 2.5$\,M$_\odot$) are found the g\,mode and p\,mode pulsators, the $\gamma$~Dor  and $\delta$~Sct stars. The theoretical instability strips for these two classes overlap. The High Amplitude Delta Scuti (HADS) stars and the older population SX\,Phe stars are subclasses of  the $\delta$~Sct class. In the same temperature range are also the high-overtone magneto-acoustic mode oblique pulsators, the rapidly oscillating Ap (roAp) stars.  Among the OB stars are the hotter $\beta$\,Cep stars ($7 \le M \le 25$\,M$_\odot$) and the cooler Slowly Pulsating B (SPB) stars\footnote{See {Section\,A.1 in the appendix} for stories of the naming of the SPB stars and discovery of the pulsation driving mechanism in B stars.} ($3 \le M \le 10$\,M$_\odot$). All of these classes have core convection zones, as a consequence of CNO-cycle H burning. A number of OBAF stars fall outside of the theoretical instability strips suggesting that the traditional pulsation classes are not pure. Nevertheless, it is convenient to use these classes as presented in Figure\,\ref{fig:hrd}, recognizing that there is overlap. 

\begin{marginnote}[]
\entry{$\gamma$~Dor stars} \  late-A to early-F main-sequence stars that pulsate in g\,modes with periods around $0.5 - 3$\,d.
\entry{$\delta$~Sct stars} \ early-A to early-F main-sequence stars that pulsate primarily in p\,modes with periods in the range 20\,min to 8\,hr; also usually pulsate in g\,modes.
\entry{roAp stars} \  Rapidly oscillating Ap stars pulsate in high-overtone magneto-acoustic modes with periods in the $4.7 - 25.8$\,min range; oblique pulsators.
\entry{SPB stars} \  B stars that pulsate in gravito-inertial modes with periods around $0.5 - 4$\,d.
\entry{$\beta$\,Cep stars} \ Early B stars that pulsate in p\,modes with periods in the range $2 - 8$\,hr; some also pulsate in gravito-inertial modes  similar to SPB stars.
\end{marginnote}

The $\delta$~Sct stars were defined traditionally as p\,mode pulsators driven by the $\kappa$-mechanism operating in the HeII ionization zone in their envelopes. The $\gamma$~Dor stars\footnote{See {Section\,A.2 in the appendix} for background on the discovery of the $\gamma$~Dor stars.} were defined as g\,mode pulsators driven by convective blocking \citep{2000ApJ...542L..57G,2005A&A...435..927D}. It is clear now that these two classes are not distinct. While there are pure g\,mode pulsators among the $\gamma$~Dor stars, there are only a few pure p\,mode pulsators among the $\delta$~Sct stars; most also pulsate in g\,modes. In addition, it can now be seen in {\it Kepler} and TESS data that many $\gamma$~Dor, $\delta$~Sct, and hotter B stars also pulsate in r~modes \citep{2018MNRAS.474.2774S}, although this is not part of the classification of these stars. \citet{2016MNRAS.457.3163X} have shown that the $\kappa$-mechanism and coupling between the pulsation and convection account for the driving and damping of both p\,modes and g\,modes in $\delta$~Sct and $\gamma$~Dor stars. 

\vspace{-5mm}

\subsection{$\delta$~Sct stars}
\label{deltasct}

These stars have a rich variety of pulsation behavior, ranging from rare singly-periodic stars, to stars with hundreds of frequencies spread across the  $0 - 80$\,d$^{-1}$ range, requiring g\,modes, r~modes, p\,modes, mixed modes, rotationally split multiplets, nonlinear combination frequencies, and possibly more to understand. It is noteworthy that these most common of all main-sequence heat engine pulsators are difficult to decipher because of the complexity of their behavior. The FTs of $\delta$~Sct stars have, until recently, defied mode identification for large numbers of their frequencies, a requisite for forward modeling. The reasons are that hundreds of modes can be excited in a given star, with radial, nonradial, and mixed modes appearing along with combination frequencies from nonlinear mode interaction. The $\delta$~Sct stars are mostly moderate to rapid rotators, so that rotationally split multiplets for nonradial modes are not equally split, and the frequency separation of the multiplets exceeds the mode frequency separation, leading to confusion in mode identification efforts. 

Further complications are caused by both amplitude and frequency modulation. \citet{2016MNRAS.460.1970B} studied 983 $\delta$~Sct stars with full 4-yr data sets from {\it Kepler}, and found more than 60\% of these stars have frequencies that vary in amplitude. Besides presenting an interesting problem of how and why some pulsation modes vary in amplitude and others do not, this study makes us aware that spurious frequencies can be generated in frequency analysis, especially when it is automated. That can produce lists of hundreds, and even thousands, of frequencies where a large fraction of them are not pulsation frequencies. Since forward modeling requires mode frequencies with mode identifications, the $\delta$~Sct stars have not yet lived up to their full potential for asteroseismology. 

Even with the large number of frequencies identified in the FTs of many $\delta$~Sct stars, pulsation models of A stars generally find an even larger number of mode frequencies excited. When only a few of those have confident mode identification, and many model frequencies are available to match, the chances of random matching increase, particularly where relatively large tolerances in the frequency matches are allowed. It is not known why so many more modes are found excited in models than are observed. \citet{2021arXiv210512553G} has recently reviewed the $\delta$~Sct and $\gamma$~Dor stars in the {\it Kepler} and TESS era, and \citet{2019MNRAS.490.4040A}, in their first-light paper on TESS sectors 1 and 2, give a broad background to the $\delta$~Sct and $\gamma$~Dor stars, to the A star zoo of metal-strong Am and Ap stars, and to the metal weak $\lambda$\,Boo and SX\,Phe stars. I refer the reader to those papers for more background and discussion of successes in advancing our understanding of $\delta$~Sct stars. I highlight two of those successes here: The discovery of asymptotic spacing of high radial overtone p\,modes in $\delta$~Sct stars (\citealt{2011Natur.477..570A}; \citealt{2020Natur.581..147B}) in this section, and the discovery of tidally forced and tidally interacting p\,modes in $\delta$~Sct stars in binaries in Section\,\ref{closebinaries}. 

A large part of the success of asteroseismology of stochastic pulsators, such as the solar-like oscillators and red giants, comes from the mode identification made possible by long series of asymptotically spaced, high radial overtone p\,modes. For decades a goal was to find such series of mode frequencies for $\delta$~Sct stars, but complications and confusion from spectral windows in the FTs of ground-based data (even multi-site data) made this difficult. The first breakthrough in finding high-overtone asymptotically spaced p\,modes in a $\delta$~Sct star came from 20\,d of {\it Kepler} data of HD\,187547 \citep{2011Natur.477..570A}. This star shows many frequencies typical of $\delta$~Sct stars, and, in addition, a series with a uniform spacing allowing a determination of $\Delta \nu$, the large separation, giving mode identification for modeling. \citet{2011Natur.477..570A} suggested that these high overtone modes could be driven by acoustic noise in the surface convection zone, even though that zone is much thinner than that in cooler solar-like stars. This idea was tested by looking at the stability of the mode frequencies in HD\,187547. For stochastically driven pulsation, the lifetimes should be short; they are only weeks to months in the Sun. As more data accumulated from the {\it Kepler} mission, \citet{2014ApJ...796..118A} found from 960-d of {\it Kepler} data that the high-overtone mode frequencies in HD\,187547 are stable -- they have lifetimes longer than 960\,d, as is evident from the sharp frequency peaks in the FT. \citeauthor{2014ApJ...796..118A} proposed a new mechanism, turbulent pressure, as the driving force of the high-overtone modes.  

A bigger breakthrough came from the discovery of 60 $\delta$~Sct stars with regular frequency spacing, giving large separations, using TESS 27-d 2-min cadence data sets and {\it Kepler} 1-min data sets of varying duration, but all at least 30\,d. \citet{2020Natur.581..147B} were able to make secure mode identifications and model these stars. They gave an example where their asteroseismic age of 150\,Myr for the young Pisces-Eridanus stellar stream supports the 120\,Myr age determined from gyrochronology of 101 low-mass members with TESS data. This newly discovered, nearby stellar stream spans 400\,pc and was previously suggested to be 1\,Gyr in age. Thus, the agreement in age from asteroseismology and gyrochronology is an important development in the study of this stellar association, and it supports the validity of both techniques. 

Asteroseismic modeling of the 60  $\delta$~Sct stars by \citeauthor{2020Natur.581..147B} also provides fundamental parameters for these stars -- mass, radius, and age -- that will have many applications: e.g.,  ages of young clusters, stellar streams, and moving groups that contain $\delta$~Sct stars,  such as the age determination of 11\,Myr for the Upper Centaurus--Lupus part of the Scorpius--Centaurus \citet[][see Section\,\ref{lambdaboo}]{2021MNRAS.502.1633M}. Knowledge of mass, radius, and age will help decipher the problem of mode selection in $\delta$~Sct stars. We do not yet know why some have hundreds of modes excited, some have only a few modes, or one mode, excited, and many stars in the instability strip do not pulsate at all. With many more $\delta$~Sct stars being observed by TESS, this form of asteroseismology is finally opening up for those most ubiquitous of upper main-sequence pulsators.

\subsection{Frequency variability in $\delta$~Sct and other pulsating stars}
\label{freqvar}

The question of frequency variability (or equivalently, period variability) in all pulsating stars is important, primarily to observe evolutionary changes in the pulsation cavities, hence to observe stellar evolution in real time. In the literature this effect is usually quantified as a rate of period change, $\dot{P}/P$. Efforts have been made to do this in most types of pulsating stars, and it is commonly invoked in white dwarf asteroseismology as a measure of cooling time (Section\,\ref{whitedwarfs}). But the effort is fraught with difficulties, calling for caution. 

Frequency variability can be driven externally by the Doppler effect for pulsating stars with companions, whether stars, planets, or even black holes. All that matters in this case is the orbital variation, and this is easily distinguished because all pulsation frequencies are affected (see {Section\,D.1 in the appendix}). 

\citet{2016MNRAS.460.1970B} studied amplitude and phase modulation of frequencies in 983 $\delta$~Sct stars observed for 4\,yr by {\it Kepler} (pulsation phase modulation is equivalent to frequency modulation). They examined causes of the amplitude and phase variations, which include unresolved frequencies (whether mode frequencies or nonlinear combination frequencies), three mode resonances, variable driving and damping, and energy exchange between coupled modes. 

Where frequency variations are detected with confidence, they can behave differently for different modes, and even have opposite signs, i.e., some frequencies increasing and others decreasing in the same star. For $\delta$~Sct stars the observed frequency variability is often orders of magnitude greater than predicted from evolutionary models, and it also often has the wrong sign\footnote{See Section A.3 in the appendix for details of the cases of 4\,CVn and KIC\,5950759.}. It is understandable evolutionarily that as mode cavities change, some modes will stretch in wavelength and others shrink to meet the boundary conditions of the cavities. Because the frequency changes may go either way, however, means that many mode frequencies in each star need to have measured frequency variability to begin to model evolutionary changes. 

We often think of stellar evolution as being a smooth process as stars move along their evolutionary tracks in the HR~Diagram. But, as with all physical processes, at some level there are probably other effects that give rise to changes that are not smooth. I think of this as ``evolutionary weather''; \citet{2017A&A...599A.116B} described it as temporary changes in the structure of the star. With the extremely high precision of frequency determinations being made with data sets that have durations of years and even decades, we now can see evolutionary weather in action.

\begin{marginnote}[]
\entry{Evolutionary weather} \  Changes in the pulsation cavity that give rise to larger frequency changes, either increasing or decreasing,  than can be explained by smooth stellar evolution.
\end{marginnote}

For frequency variability that is obviously not evolutionary, the physical causes are not known. This is a case in asteroseismology where the space data do not always dominate the studies, since a long time base is useful, and for small frequency changes, is needed. Note, e.g., the 160-yr study of Polaris, one of the best-known stars in the sky. Polaris is a Cepheid that pulsates in the fundamental radial mode. \citet{2005PASP..117..207T} examined data obtained from 1844 -- 2004. They found a period change of $\sim$4.5\,s\,yr$^{-1}$, consistent with evolution during a redward first crossing of the instability strip. But they also found that the rate of change had a rapid decrease during 1963 -- 1966, after which it continued as before. That was coupled with a steady pulsation amplitude for Polaris until 1963 -- 1966, after which the amplitude was much smaller and somewhat erratic. This is what I call evolutionary weather. It gives us pause when interpreting pulsation frequency changes as evolutionary.

\subsection*{Summary: The $\delta$~Sct stars}
\begin{enumerate}
\item $\delta$~Sct stars are the most common heat-engine main-sequence pulsators. 
\item Hundreds of frequencies can be excited, and in models even more are predicted. 
\item The mode selection mechanisms are not known.
\item Recent determinations of the large separations in many $\delta$~Sct stars have opened up asteroseismic inference for these stars: mass, radius, and age.
\item Precise asteroseismic ages agree well with gyrochronology, validating both methods.
\end{enumerate}

\subsection{g\,mode pulsators: $\gamma$~Dor, $\delta$~Sct and SPB stars}
\label{gmodepulsators}

Observable g\,modes in upper main-sequence stars are of fundamental importance because these modes have pulsation cavities that sample the conditions in the deep interior of the stars. It is difficult to detect g\,modes in cooler stochastic pulsators, such as the Sun, because they cannot propagate in the thick outer convection zones, so that little of the signal reaches the surface. It was the advent of CoRoT, then {\it Kepler} and TESS that led to an explosion of discovery -- both observational and theoretical -- from g\,modes. 

We did not realize before these space missions, although it is in hindsight obvious: the g\,mode frequencies are so closely spaced that typically 90\,d or more of observations are needed to resolve them. Over that time span gaps in ground-based observations cause a plague of aliases and  tangled spectral window patterns in the FTs, so that the nearly continuous data of the space missions are required for studying these stars. It is the 4-yr data sets of {\it Kepler} and the 1-yr continuous viewing zone TESS data that have opened new fields of research with g\,modes in $\gamma$~Dor, $\delta$~Sct, and SPB stars. 

The new discoveries are the observations of deep interior rotation, allowing inference of angular momentum transport, of measurement of core overshoot at the convective core boundary, and of measurement of the mass and radius of the convective core, giving evidence of chemical mixing and extended stellar lifetimes. See \citet{2019ARA&A..57...35A} for an extensive review of our new knowledge of angular momentum transport in stars.

The exploitation of the long duration space data sets for $\gamma$~Dor stars is one of the outstanding achievements of asteroseismology in recent years. It has given us an observational view of these stars that could not have been obtained from the ground, and it has driven a plethora of theoretical developments to extract astrophysical inference from those data. In a study of 611 $\gamma$~Dor stars with 4-yr {\it Kepler} data sets, \citet{2020MNRAS.491.3586L} found that essentially all $\gamma$~Dor stars pulsate in dipole g\,modes, 30\% also pulsate in quadrupole g\,modes, and 16\% in addition pulsate in r~modes. The most visible modes are the prograde sectoral dipole modes ($\ell =1, m=1$). Those often have many consecutive radial overtones for which the frequencies are easily identifiable in an FT. As was first derived by \citet[equation 23]{1979PASJ...31...87S}, in the asymptotic limit for g\,modes in a nonrotating, nonmagnetic star, the pulsation periods are roughly uniformly spaced by an amount that depends on the characteristic period spacing, $\Pi_0$, and mode degree, $\ell$.  Retrograde dipole g\,modes are also excited, whereas zonal dipole modes are less commonly seen. 

These series of g\,mode overtones greatly facilitate mode identification. We now have new tools to measure internal rotation, and even differential rotation from core to surface, leading to the ability to calculate internal angular momentum transfer, and to measure the core mass, radius and amount of convective overshoot, which affect mixing in the energy generating layers, hence main-sequence lifetimes. These are major advances improving stellar structure and evolution models.

\subsection{Core to surface rotation from multiplets}
\label{coretosurface}

\begin{figure*}[t]
\centering
\includegraphics[width=0.8\linewidth,angle=-90]{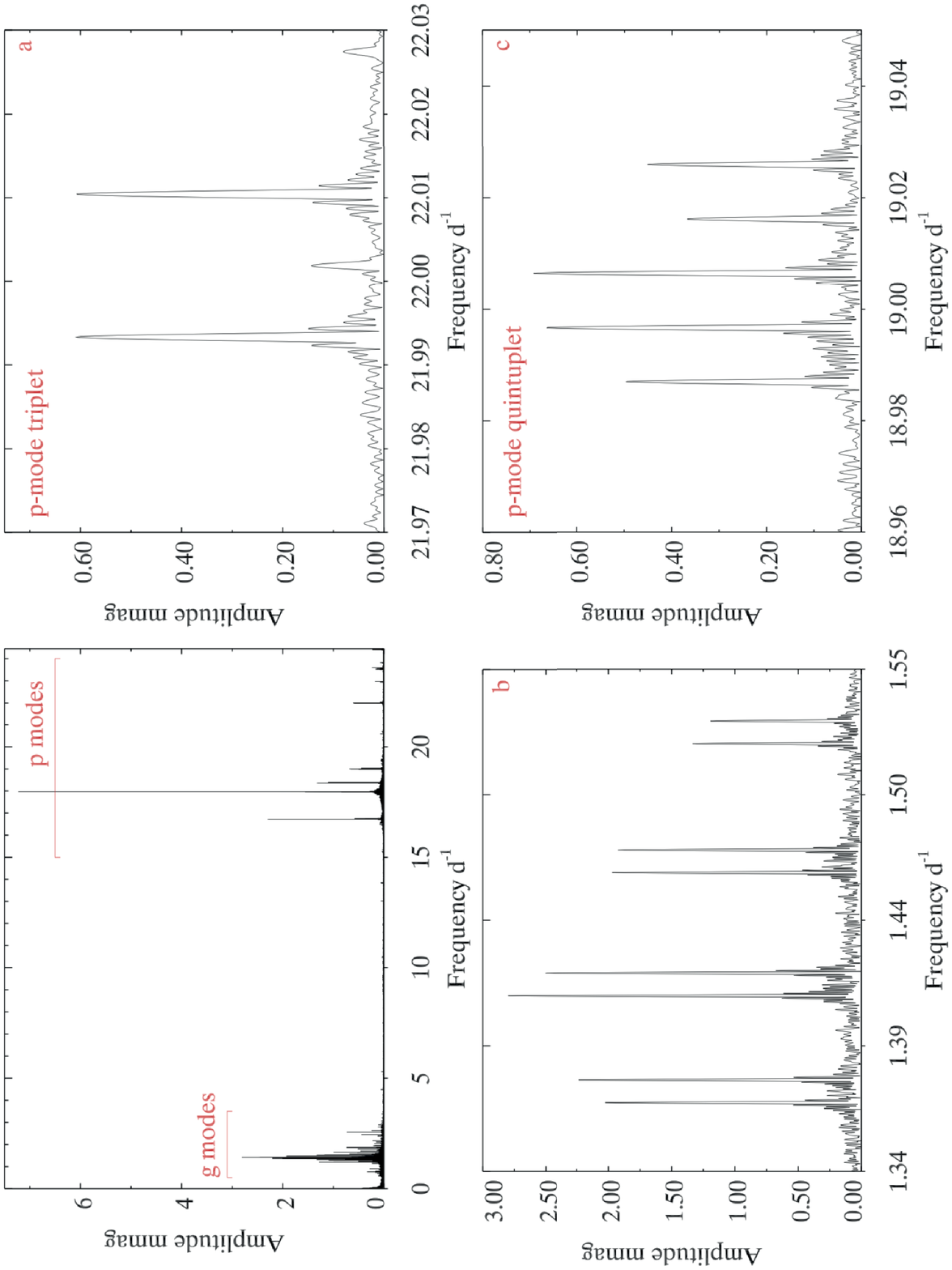}
\vspace{-1.5cm}
\caption{The top left panel shows the FT for 4 yr of {\it Kepler} data for the A6 main-sequence $\delta$~Sct -- $\gamma$~Dor star KIC\,1145123, for which near-core to surface rotation was measured for the first time in a hydrogen-burning main-sequence star. The frequency ranges of the g~modes and p~modes are marked. The bottom left panel shows an expanded view for the g~modes where four of many consecutive overtones can be seen to be split into apparent doublets. Those  are actually dipole rotational triplets with the central peak not visible at this scale. The two right panels show a dipole rotational triplet and quadrupole rotational quintuplet for p~modes. The asteroseismologist's tasks here are the recognition of the rotational multiplets, the g~mode series, and the identification of the overtones of the p~modes by forward modeling. This star has a particularly clear view of the patterns as a consequence of its very slow rotation ($P_{\rm rot} \sim 100$\,d). These figures are from Figures 1 and 3 of \citet[][``Asteroseismic measurement of surface-to-core rotation in a main-sequence A star, KIC\,11145123'']{{2014MNRAS.444..102K}}.}
\label{fig:11145123}
\end{figure*}

KIC\,11145123 is a $\delta$~Sct star with stunningly simple rotational multiplet patterns of dipole and quadrupole modes for both g\,modes and p\,modes that provided the first direct measure of both internal and surface rotation in a H-core-burning star \citep{2014MNRAS.444..102K}. Figure\,\ref{fig:11145123} shows an FT for this star with higher resolution views of the g~mode and p~mode rotational multiplets. Measuring stellar rotation from frequency splitting of dipole and quadrupole frequency multiplets depends on the Ledoux constant, $C_{n,\ell}$ (e.g., see \citealt[eqns $44-46$]{2021RvMP...93a5001A}), which asymptotically approaches 0.5 for high radial overtone dipole g\,modes, and has a small value not much greater than zero for p\,modes. As a consequence, \citeauthor{2014MNRAS.444..102K} derived for KIC\,11145123 a model-independent, near-core rotation period of 105.1\,d, and a surface rotation period of 98.5\,d. The surface rotates faster than the deep interior -- a surprise at the time. 

Simple arguments from stellar evolution and conservation of angular momentum lead to the expectation of the shrinking core spinning up and the expanding envelope spinning down with age. Clearly, a new view of interior angular momentum transfer in main-sequence stars, as well as angular momentum gain from accretion, or loss from mass loss, had become possible. This field has developed dramatically with interior rotation being observed in nearly 2000 stars from the main-sequence through red giants and on to white dwarfs. See \citet{2019ARA&A..57...35A} for a theoretical review and \citet[figure 6]{2021RvMP...93a5001A} for an observational view of interior rotation versus $\log g$, hence evolutionary stage.

Examining KIC\,11145123 further, \citet{2019ApJ...871..135H} used a detailed perturbative approach to obtain a 2D view of the rotation in this star. By studying mixed modes, which have some sensitivity to the convective core rotation, they derived a core rotation rate for $r/R < 0.05$ that is six times faster than the radiative exterior. Thus there is a strong velocity shear at the convective core boundary that could impact mixing there, hence main-sequence lifetime. They also found marginally significant evidence of surface differential rotation, with higher latitudes rotating faster than the equatorial region, by exploiting the difference in the latitudinal dependence of dipole and quadrupole modes. \citet{2016SciA....2E1777G} also used dipole and quadrupole p\,modes to calculate both polar and equatorial radii for KIC\,11145123, remarkably finding the difference to be only $3 \pm 1$\,km. This star is so round that some unknown factor is needed to suppress rotational oblateness, even at its very slow rotation rate. 

Other very slowly rotating stars were then found in the {\it Kepler} data and analyzed. \citet{2015MNRAS.447.3264S} found KIC\,9244992 to be similar to KIC\,11145123, with an interior rotation period of 64\,d and a surface rotation period of 66\,d; at least the surface is rotating more slowly than the interior in this case. Then \citet{2015ApJ...810...16T} found a core boundary rotation period of 71\,d for the SPB star KIC\,10526294, and suggested that the seismic data supported a counter-rotating envelope! That is amazing, if correct.

For $\delta$~Sct stars where both p\,modes and g\,modes are observed, both deep interior and surface rotation can be measured from the multiplet splitting in the FTs. For pure $\gamma$~Dor stars where only g\,modes are observed, it is still possible to examine core and surface rotation rates by deriving the core rate from the period spacing of the g\,modes and constraining the surface rate from a determination of the radius and spectroscopically measured $v \sin i$. \citet{2016MNRAS.459.1201M} showed this for KIC\,7661054 which has a core rotation period of 27\,d. 

But these few stars discussed here are all very slow rotators. Most A and early-F main-sequence stars rotate much more quickly, with typical rotation periods of a day or two. For g\,modes in these stars the rotation period and pulsation periods are similar. To observe interior rotation and infer angular momentum transfer in upper main-sequence stars, new techniques and new theoretical developments were needed. This has resulted in an explosive outpouring of research, both observational and theoretical, in this new field. 

\subsection{Internal rotation from g\,modes and r~modes in upper main-sequence stars}
\label{internalrotation}

A big breakthrough in determining internal rotation in $\gamma$~Dor and SPB stars came when it was shown that for a series of g\,modes with high radial overtones, the difference in the periods, $\Delta P$, for consecutive overtones versus the periods, $P$, provides a direct measure of rotation in the pulsation cavity, with most sensitivity just above the convective core deep in the star. \citet{2013MNRAS.429.2500B}, \citet{2016A&A...593A.120V}, and \citet{2017MNRAS.465.2294O} showed how the slope of points in the $\Delta P - P$ diagram measures the interior rotation rate. The slopes differ for series of prograde sectoral, zonal, and retrograde sectoral dipole modes \citep[figure 2]{2017MNRAS.465.2294O}, and all series are useful. The prograde modes are most common and in many cases dozens of consecutive radial overtone modes are detected, particularly in the 4-yr {\it Kepler} data where the frequency resolution is the best to date. \citet[][figure 5]{2019A&A...626A.121O} showed the good agreement between the internal rotation rates determined by \citet{2018A&A...618A..47C} and by \citet{2016A&A...593A.120V}, supporting the robustness of the two methods.

This new possibility was exploited immediately.  \citet{2017ApJ...847L...7A} studied the internal rotation of 59 $\delta$~Sct/$\gamma$~Dor stars plus 8 SPB stars covering a mass range of $1.4 - 5$\,M$_\odot$ for which they also had constraints on surface rotation from spectroscopic $v \sin i$. For rotation velocities up to 50\% of breakup, they found that the stars deviate from solid body rotation only mildly. With comparison results for red giants, they concluded that core rotation rates must drop significantly in the evolutionary interval where the stars shift  from H to He core burning. It is clear that efficient mechanisms for the transport of interior angular momentum during the evolution of stars must be operational. 

One proposed mechanism is Internal Gravity Waves (IGW), which are discussed in detail by \citet{2013ApJ...772...21R} and \citet{2019ARA&A..57...35A}. (IGW should not be confused with gravity modes.) The IGW are generated by turbulence at the convective core boundary. As they travel out into the radiative envelope, where density and temperature are dropping, they can become shock waves, break and deposit energy and angular momentum, and contribute to chemical mixing.  

\citet{2021A&A...646A..19T} developed a formalism to model the evolution of a 1.5-M$_\odot$ star including the coupled effects of a global magnetic field and rotation. Their model settled to near-rigid rotation on the Alfv\'en timescale. In another model, angular momentum transport arises from magnetic torque caused by the Tayler instability \citep{2019MNRAS.485.3661F}. Interestingly, \citet{2019ApJ...881L...1F} examined that angular momentum transport mechanism and concluded that most single black holes, as well as many binary black holes, are born rotating slowly, with implications for future gravitational wave detections. Thus asteroseismology even sheds light on gravitational wave astronomy.

Two new techniques for determining internal rotation have been developed that have the potential to be applied to many more $\gamma$~Dor, $\delta$~Sct, and SPB stars across a range of rotation rates. These are the $\nu - \Delta \nu$ technique of \citet{2020A&A...635A.106T}, and the exploitation of dips in the $\Delta P - P$ diagram by \citet{2020A&A...640A..49O} and \citet{2021MNRAS.502.5856S}. Details for these techniques are discussed in {Section\,A.4 in the appendix}. 

\subsection{Probing the convective core boundary}

For many years, a goal in studying the SPB and $\beta$\,Cep stars has been to determine the extent of core overshoot -- which affects mixing and main-sequence lifetime -- and core rotation. Our crude understanding of the extent of core overshoot is one of the most uncertain parameters in stellar models. See \citet{2013PhDT.........6P} for a historical review of the observational efforts. Until recently, the number of OB stars for which core overshoot or core rotation could be determined was limited to only 11 stars \citep{2013EAS....64..323A}. That is now changing with analysis of {\it Kepler} data. TESS data in the 1-yr continuous view zone also have sufficient duration. These studies will come to fruition with analysis of the TESS Full Frame Images (FFI), which were originally 30-min cadence and are now 10-min cadence. 
 
 The $\beta$\,Cep stars are rarer than the SPB stars as a consequence of the stellar initial mass function and main-sequence lifetimes. Since all of them have strong wind losses, and some of them are potential progenitors of core-collapse supernovae, they are of great interest in understanding their contribution to galactic chemical evolution. However, because of their scarcity, up to now only 12 of these stars have been modeled asteroseismically. \citet{2019MNRAS.489.1304B} discussed some $\beta$\,Cep stars discovered in {\it Kepler} K2 data, and pointed out the asteroseismic potential of these stars once spectroscopic observations provide mode identifications. I refer the reader to the comprehensive reviews and discussions of OB stars and their pulsations by \citet{2020FrASS...7...70B, 2020svos.conf...53B}, who provides extensive background on asteroseismology and discussion of ground-based and space photometry of these stars and their potential.  

As a consequence of core nuclear burning, there is a strong gradient in the mean molecular weight -- a $\mu$-gradient -- at the boundary of the convective core in upper main-sequence stars. That boundary is a turning point for g\,modes; for each radial overtone g\,mode the standing wave must have a node at the boundary. As the radial overtone increases, the radial wavelength of the mode decreases while the mode cavity remains fixed. The mode wavelength adjusts to give an integral number of radial waves in the cavity, and the frequency adjusts with that. This produces cyclic dips in the $\Delta P - P$ diagram superposed on the rotational gradient. \citet{2008MNRAS.386.1487M} first showed how these cyclic dips can be used to measure the radius of the convective core. That depends on core overshoot and mixing at the core boundary, since the introduction of fresh H fuel to depths where the temperature is high enough for CNO-cycle fusion modifies the convective core radius. That fresh fuel also extends the main-sequence lifetime, so knowledge about core mixing is important in stellar evolution models. The cyclic behavior on the $\Delta P - P$ diagrams is evident in both calculated and observed diagrams (\citealt{2013MNRAS.429.2500B}, \citealt{2016A&A...593A.120V}, \citealt{2017MNRAS.465.2294O}). 

This oscillatory character is also present in the period spacings of g\,modes in white dwarfs and has been used for decades to measure the extent of the surface layers that are the pulsation cavities in those stars. It is now used to determine the position of the $\mu$-gradient boundary in $\gamma$~Dor and SPB stars. \citet{2019MNRAS.485.3248M} combined asteroseismology and spectroscopy to estimate the core masses and convective overshooting in $\gamma$~Dor stars to conclude that efficient angular momentum transport is already active for main-sequence stars. 

A breakthrough in modeling SPB stars has come from \citet{2021NatAs...5..715P} who selected 26 SPB stars out of a sample of 60 with {\it Kepler} long cadence 30-min 4-yr data sets. They obtained spectra for 15 of these stars, giving boundary conditions of surface temperature and gravity, metallicity, and projected equatorial rotation velocity; Gaia astrometric data provided luminosities. With a grid of models with free parameters that include mixing at the core envelope boundary and internal rotation, \citeauthor{2021NatAs...5..715P} modeled the g\,mode period spacings. They examined a number of envelope mixing sources, including convective envelope penetration (core overshoot), internal gravity waves (IGW), vertical shear and meridional circulation combined with vertical shear. The stars have a range of internal rotation frequencies up to 70\% of the critical breakup frequency for 9 of them. Results show the fractional core mass to be 30\% near the zero-age main-sequence and only dropping to 6\% (not zero) at the terminal-age main-sequence where core H fusion finally ceases. They found internal mixing profiles of SPB stars that are radially stratified instead of constant. 

All of these results are a major step in modeling stellar evolution of upper main-sequence stars, with asteroseismic observations of interior conditions from core to surface.  \citet{2021A&A...646A.133H} discussed their theoretical 2D hydrodynamical simulations to understand the mixing that occurs between the convective core and radiative layers above it in $1.3 - 3.5$\,M$_\odot$ stars. The results of \citet{2021NatAs...5..715P} and future expansion of these observational asteroseismic studies will allow constraints on the development of 3-D hydrodynamical simulations. 

\subsection*{Summary: Internal rotation and the convective core boundary}
\begin{enumerate}
\item Prior to the photometric space missions internal rotation was known essentially only for the Sun, and that was found to be completely different to what was predicted prior to helioseismology.
\item We now measure the internal rotation profiles in thousands of stars from the main-sequence to white dwarfs, allowing direct observation of angular momentum transport with stellar evolution. 
\item Strong angular momentum transport mechanisms are needed in stars.
\item Asteroseismology measures the size of the convective core with impact on asteroseismic age determinations. 
\end{enumerate}

\subsection{Atomic diffusion}
\label{diffusion}

Atomic diffusion -- radiative levitation and gravitational settling -- has impact on stellar structure and evolution across the HR~Diagram. It is a small component of the standard model of the Sun, it significantly affects globular cluster ages determined from model isochrones, it is active in white dwarf atmosphere stratification, and it produces the atmospheric abundance anomalies of the metallic-lined Am and peculiar Ap stars, the most peculiar stars known (Section\,\ref{peculiar}). See \citet{2020pase.conf..185M} for a conference review of the current status of atomic diffusion for Ap stars, and \citet{2015ads..book.....M} for a complete treatise on the subject. See {Section\,A.5 in the appendix} for two case examples in asteroseismic models. 

\subsection{r~modes}
\label{rmodes}

\begin{figure*}[t]
\centering
\includegraphics[width=1.0\linewidth,angle=0]{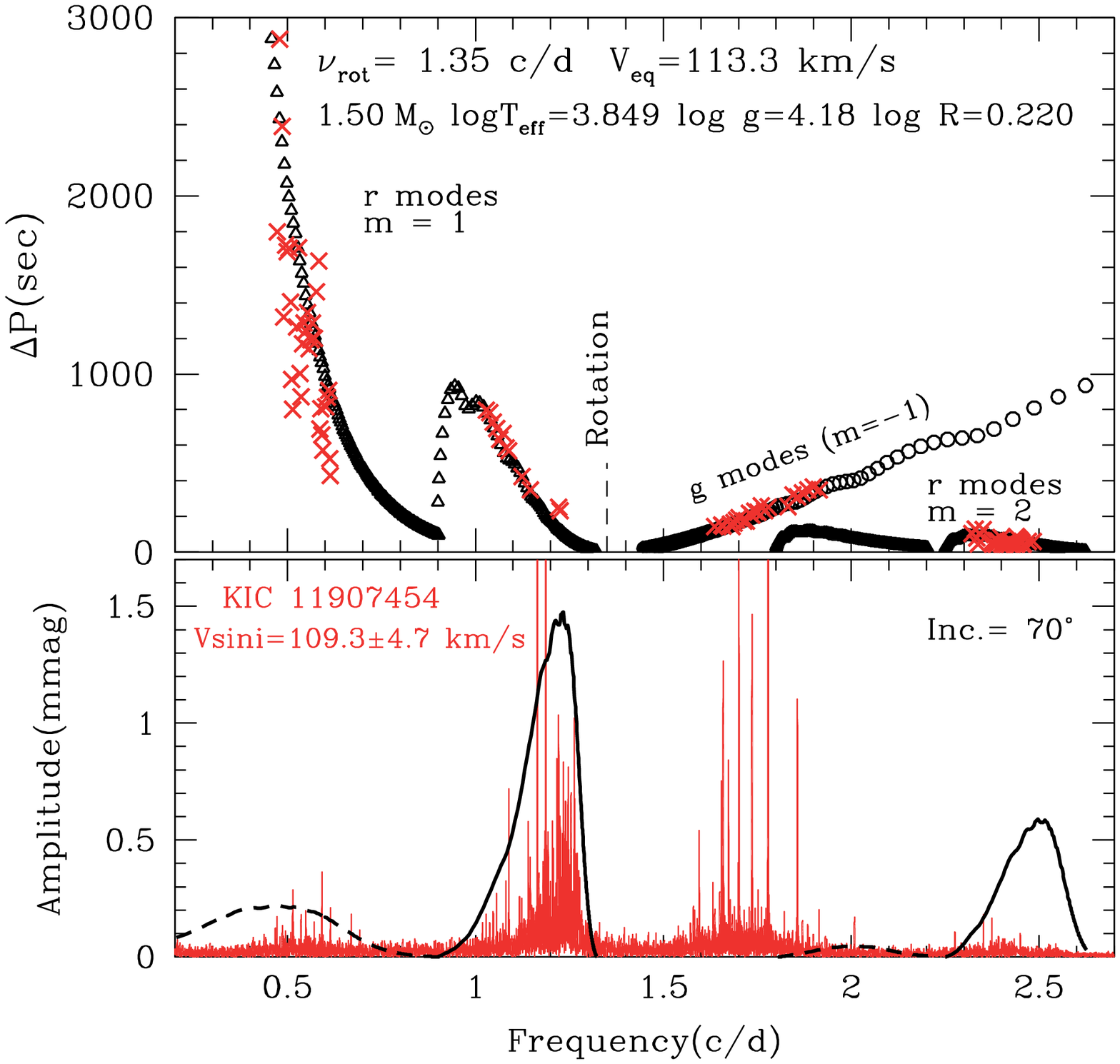}
\caption{Observed period spacings (crosses) and an FT for the early-F $\gamma$~Dor star KIC\,11907454 are compared with predicted period spacings (triangles and circles) of r and g~modes (top panel) and the expected amplitude distribution of r~modes for an inclination angle of $70^\circ$ (bottom panel). Black solid  lines (for even modes) and dashed lines (for odd modes) show predicted amplitude distributions normalized to fit the hump. The adopted rotation frequency is indicated by a vertical dashed line (top panel), which agrees with  the observed rotation frequency $1.35\pm0.02$~d$^{-1}$ obtained for KIC\,11907454 by \citet{2016A&A...593A.120V}. This diagram shows the dense spacing of the r~modes and g~modes, highlighting the need for months to years of continuous light curves to decipher the FTs, data only available from space missions. 
This figure and caption are adapted from Figure 5 of \citet[][``Theory and evidence of global Rossby waves in upper main-sequence stars: r-mode oscillations in many Kepler stars'']{2018MNRAS.474.2774S}.}
\label{fig:rmodes}
\end{figure*}

Global Rossby waves, r~modes, are inertial modes that are now observed in upper main-sequence rotating stars, where the toroidal motion of the r~modes couples with the spheroidal motion of the Coriolis force causing temperature variations that lead to visible light variations. The r~modes are retrograde in the co-rotating frame, and they have mathematically an infinite set of eigenmodes, as do p\,modes and g\,modes. The r~modes have frequencies that are so closely spaced that even with {\it Kepler} 4-yr data sets they are not always fully resolved, hence what is seen in the FT is a hump of amplitude with partially resolved peaks. Figure\,\ref{fig:rmodes} shows an example of an FT and period spacing for a $\gamma$~Dor star, where the dense hump of closely-spaced r~mode frequencies is evident, as well as a series of g-mode frequencies. The model provides an acceptable match to the observations, confirming the r~mode interpretation of the frequency humps.

\citet{2016A&A...593A.120V} first recognized the signature of r~modes in FTs of 10 $\gamma$~Dor stars. \citet{2018MNRAS.474.2774S} then presented the theory of r~modes, modeled the 10 stars studied by \citeauthor{2016A&A...593A.120V}, and noted the signature of r~modes in many stars with {\it Kepler} data, including $\gamma$~Dor, $\delta$~Sct, SPB stars, an outbursting B emission-line (Be) star (Section\,\ref{bestars}), and binary heartbeat stars (Section\,\ref{hbstars}). Thus r~modes are ubiquitous across the upper main-sequence. 

\citet{2018MNRAS.474.2774S} found the main low-frequency hump in the FT to be from $m = 1$ r~modes with frequencies just below the rotation frequency of the star. Often there is a sharper, generally unresolved peak at a slightly higher frequency than the hump, which they attributed to rotational variation caused by a faint spot or spots. The photometric amplitude for these spot variations is usually of order 10 ppm, hence the spots have much less contrast than those in the Sun where the ensemble of spots creates rotational variation up to 1 ppt. \citeauthor{2018MNRAS.474.2774S} suggested the driving for the r~modes comes from g\,modes, from deviated flows past spots, from mass outbursts (as in Be stars) and/or from tidal forces (as in heartbeat stars, Section\,\ref{hbstars}). Some surface variation, such as spots, is needed for visibility, and possibly for driving, of the r~modes. See {Section\,A.6 in the appendix} for a short discussion of evidence for weak spots in upper main-sequence stars. 

Most $\gamma$~Dor stars do not show the long series of sectoral or zonal modes needed for a $\Delta P - P$ diagram. The mode selection mechanisms in these stars are not understood, and it is also not known whether the stars that do show mode series for plotting $\Delta P - P$ diagrams are the same in their internal rotation as the majority of these stars that do not show such series. \citet{2019MNRAS.487..782L} inspected 1593 {\it Kepler} stars in the temperature range $6600 - 10\,000$\,K, which encompasses the range of the $\delta$~Sct and $\gamma$~Dor stars, finding 82 that have resolved period spacing patterns for both g\,modes and  r~modes, the largest sample yet. All of those stars are either in the $\gamma$~Dor instability strip, or only somewhat hotter than that. \citeauthor{2019MNRAS.487..782L} determined near-core rotation rates and the characteristic periods, $\Pi_0$, for the 82 stars, noting that the g\,modes typically have radial overtones centered around $n \sim 55$, whereas the r~modes have lower radial overtones in the $13 \le n \le 50$ range. The near-core rotation periods were in the range $0.4 - 2.5$\,d. Using $\Pi_0$ is an indicator of stellar age, since it decreases over the main-sequence lifetimes of $\gamma$~Dor stars, they found no correlation between internal rotation and age, although that could be because of too small a range in ages.  For 5 out of 6 stars where they could detect a signature of surface rotation, the stars have uniform rotation. 

\subsection*{Summary: r~modes}
\begin{enumerate}
\item r~modes are common in upper main-sequence stars.
\item They provide a good measure of the surface rotation period.
\end{enumerate}

\subsection{Asteroseismology at the top of the HR~Diagram}
\label{slf}

Until very recently, asteroseismology of the most luminous blue stars was limited by a dearth of stable pulsation modes. Wolf-Rayet stars and Luminous Blue Variables (LBV), such as the iconic $\eta$\,Carinae, S\,Doradus, and P\,Cygni, have variability that is generally not periodic, especially during outburst, although some show pulsations in quiescence. There are also the  $\alpha$\,Cyg stars, which are supergiant B and A stars that pulsate with periods of days to weeks. The long periods have made it difficult to obtain sufficient data for asteroseismology of these stars. 

These massive stars form near the Eddington limit where where their fierce radiation pressure puts an upper limit to the greatest mass a star can have. They are the precursors of core-collapse supernovae, hence of neutron stars, black holes, gamma ray bursters, and the sources of detectable gravitational wave events. Their evolution is complex, with mass loss rates, internal angular momentum transport, and mixing all uncertain. Models show blue loops, so that stars may be crossing the HR~Diagram from blue to red, or vice versa, making it challenging to distinguish their evolutionary state \citep{2021A&A...650A.128G}.

\begin{marginnote}[]
\entry{LBV} \  Luminous Blue Variables are hot supergiants that generally show irregular light variations. 
\entry{$\alpha$\,Cyg stars} \  BA supergiants showing nonradial pulsation with periods of weeks to months. Some LBV show periodic $\alpha$\,Cyg variations when in quiescence.
\entry{Strange modes} \  Non-adiabatic modes trapped in thin surface layers by a density inversion in high $L/M$ stars. There is no one-to-one correspondence between radial overtone, $n$, and mode frequency.
\entry{SLF variables} \  Stochastic Low Frequency variables are main-sequence and supergiant stars that show stochastically excited low frequency variability.\end{marginnote}

Stars near the Eddington limit, including both massive stars and hydrogen-deficient eHe stars (Section\,\ref{pvtel}), have high luminosity to mass ratios. For these stars highly non-adiabatic, opacity-driven pulsation modes with periods in the $10-100$\,d range may be confined to a layer in the outer envelope where a density inversion provides the lower boundary to the thin pulsation cavity. In these cases, there is not a simple correspondence between the mode frequency and the radial overtone. These are the strange modes \citep[see, e.g.,][]{1998MNRAS.294..622S, 2016MNRAS.458.1352J}. As seen in Figure\,1, they occur across a number of pulsation classes, including the LBV,  $\alpha$\,Cyg, WR, PV\,Tel, and R\,CBr stars. It has been proposed that strange mode amplitudes may reach escape velocity and be implicated in LBV winds and outbursts \citep{1999MNRAS.303..116G}, interestingly even for primordial, metal-free, supermassive stars at the beginning of the chemical evolution of the universe \citep{2018MNRAS.475.4881Y}.

A related phenomenon to strange modes is surface trapped modes where the hydrogen partial ionization zone, rather than a density inversion, provides the lower pulsation cavity boundary. These have been found in models of RV\,Tau, BL\,Her, and W\,Vir Cepheids in different crossings of the instability strip \citep{2016MNRAS.456.3475S}. It is also possible that some strange modes are driven by convection \citep{1981PASJ...33..427S}. Problems with mode identification, low numbers of modes, long periods, and mode instabilities have not been conducive to asteroseismology of strange modes yet. 

However, now many of the most massive stars are amenable to asteroseismology. \citet{2019NatAs...3..760B,2020A&A...640A..36B} announced the discovery of Stochastic Low Frequencies (SLF) in essentially all OB stars from the main-sequence to supergiants. SLF are correlated with stellar mass and age, and with surface macroturbulence. With SLF by themselves, and combined with cases where there are coherent mode frequencies, this opens up asteroseismology for these rare, important, massive stars. The TESS mission is providing a wealth of observations of these stars. 

\subsection{\ Pulsating Be stars}
\label{bestars}

The Be stars range from O to early-A and all, by definition, show emission lines in their spectra\footnote{See {Section\,A.7 in the appendix} for the early history of Be stars.}. They are also all pulsating, primarily in g\,modes, and many show aperiodic outbursts that are detected both spectroscopically and photometrically. \citet{2020arXiv201013905L} provide a review of Be stars characteristics, along with an analysis of  432 Be stars from the first year -- the southern ecliptic hemisphere -- of TESS data. They found that nearly all have low frequency variability typical of SPB stars, i.e. g\,modes, that 17\% show frequencies in the range of p\,modes, and that many show evidence of r~modes. We can now also expect SLF variations to be found in the hotter Be stars. They rotate with equatorial velocities that are greater than 80\% of break-up velocity. Pulsation and rapid rotation are fundamental physical characteristics of these stars. During that first year of TESS data 17\% of the 432 Be stars had photometric outbursts -- brightening events -- and those outbursts, whether detected photometrically or spectroscopically, are another fundamental character of Be stars, although not exclusive to them. \citet{2021A&A...655A..59V} found that a fraction of 38 SPB stars that they studied for nonlinear interactions show outbursts -- without being Be stars. 

\begin{marginnote}[]
\entry{Be stars} \  Pulsating, rapidly rotating O to early-A stars that have outbursts and show emission lines from a decretion disk.
\end{marginnote}

Be stars have decretion disks -- circumstellar disks that are supplied with material from equatorial mass loss during outbursts. There is no agreement about the physical mechanism for the mass loss, but a leading proposal is that horizontal prograde g\,mode pulsation velocities from many modes can simultaneously add to the near break-up rotational velocities to give equatorial lift-off of material during outburst to populate the decretion disk, hence provide gas in that disk to create the characteristic emission lines -- first seen by \citet{1866AN.....68...63S}. TESS is observing almost all known Be stars. Asteroseismology promises to fill out our understanding of interior rotation and mixing in these stars and illuminate further the outburst mechanism. 

\subsection{\ Nonlinear pulsation: Combination frequencies -- opportunities and pitfalls}
\label{combinations}

Photometric time series -- light curves --  are the principal means by which we obtain pulsation mode frequencies for asteroseismic modeling. What we are observing is a change in the flux from the observed hemisphere, usually construed to be a change in luminosity, although that is not always the case, as for the tidally tilted pulsators  (section\,\ref{ttp}). The observed flux differences depend on changes in $T_{\rm eff}$ and changes in cross-section. The latter depends directly on changes in $R$ for radial modes, but may be more or less complex for nonradial modes. For example, dipole modes show no change in cross-section during pulsation, hence light variations from them are purely the result of temperature changes. 

For low amplitude pulsators, where the pulsation is only a perturbation to the structure of the star, to first order the luminosity variations can be linearized as a function of $T_{\rm eff}$. Nevertheless, a sinusoidal variation in $T_{\rm eff}$ does not necessarily produce a sinusoidal light curve. Stars with multiple pulsation modes also experience coupling among the modes. Changes in the mode cavity of one mode affect the pulsation of other modes with different cavities. This mode coupling also produces non-sinusoidal light curves. Because we are thinking in terms of Fourier analysis when we extract pulsation mode frequencies, we naturally find in the FT frequencies that are harmonics and combinations of the actual mode frequencies. Those are the Fourier description of the non-sinusoidal variations of the light curve. Since asteroseismology models pulsation mode frequencies, it is imperative to identify the harmonics and combination frequencies so as not to match them to model frequencies in forward modeling, unless, of course, they should resonantly excite a mode frequency. 

On the other hand, it is possible to use the combination frequencies, i.e., the non-sinusoidal shape of the light curve, for asteroseismic inference. \citet{2005ApJ...633.1142M}, building on the work of \citet{2001MNRAS.323..248W} and \citet{1992MNRAS.259..519B}, showed how the non-sinusoidal  light curves of white dwarf stars can be used to infer the thermal response timescale of the convection zone, hence its depth. He pointed out that the technique has application to Cepheids and RR\,Lyr stars. See \citet[section III.A.2]{2021RvMP...93a5001A} for further discussion and applications. 

Observationally, we must recognize the combination frequencies. That seems straightforward, but there have been many stumbles in previous studies. \citet{2015MNRAS.450.3015K} examined the frequency groups often seen in FTs of $\gamma$~Dor, SPB, and pulsating Be stars. Those frequency groups had previous been  associated solely with rotation, and it is the case that g\,modes and r~modes can produce frequency groups associated with rotation \citep{2018MNRAS.474.2774S}. But what \citeauthor{2015MNRAS.450.3015K} showed for the extreme cases of stars with frequency groups is that there are only a few pulsation mode parent frequencies and the rest of the many peaks that dominate the FTs are harmonics and combinations as a result of non-linear interactions of the parent modes. They showed, both observationally and theoretically, that combination frequencies can have higher observed amplitudes than parent mode frequencies, calling for even greater care in identifying mode frequencies for asteroseismic modeling. They also found that combination frequencies in a Be star, both in outburst and between outbursts, support the interpretation of the star as a rapidly rotating SPB star. \citet{2021A&A...655A..59V} make a plausible case that Be stars are complex SPB stars at the end of a range of outburst behavior for SPB stars. The two classes are not independent.

The technique for identifying combination frequencies has been to identify (or often assume) parent mode frequencies, then simply calculate the combination frequencies and remove them from the FT by pre-whitening, as well as disregard them in forward modeling. It is important to remove the combination frequencies during frequency analysis, since they can have high amplitudes (or even have the highest amplitudes) and their spectral windows can hide lower amplitude mode frequencies. \citet{2020MNRAS.498.1194L} recognized this and developed a self-consistent method to extract combination frequencies. \citet{2021MNRAS.504.4039B} showed a plethora of combination frequencies and harmonics in the $\delta$~Sct star KIC~5950759, which \citet{2020MNRAS.498.1194L} used to illustrate their method.

\subsection{\ Chemically peculiar stars and pulsation: Am, Ap, and $\lambda$\,Boo stars} 
\label{peculiar}

The B, A, and F stars on the upper main-sequence are home to a zoo of spectroscopically peculiar stars; see the introduction of \citet{2000BaltA...9..253K} for a detailed guide to the numerous classes and arcane naming conventions. The three principal types I discuss here are the metallic-lined A (Am) stars, the magnetic peculiar A (Ap) stars, and the metal-weak $\lambda$\,Boo stars. 

\subsubsection{Am and Ap stars}
\label{AmandAp}

The Am and Ap stars have surface abundance anomalies that can reach overabundances of some rare earth elements of a stunning factor of $10^6$ in comparison with the Sun, while a few other elements or ions are deficient. The physical mechanism,  atomic diffusion (Section\,\ref{diffusion}), is most readily observable in the Am and Ap stars, hence their wider interest in stellar astronomy. The Am and Ap stars are slow rotators, generally with $v \sin i \le 100$\,km~s$^{-1}$, a necessary condition for atomic diffusion to produce observable anomalies in the atmosphere, as more rapid rotation generates faster meridional flows that become turbulent, keeping the outer layers mixed. The Am stars have no, or very weak ($\sim$1\,G) magnetic fields, whereas a subclass of the Ap stars, the magnetic Ap stars, have global magnetic fields with strengths of kiloGauss.  The Ap stars constitute about 10\% of main-sequence A stars, whereas the Am stars are much more common, reaching a peak of about 50\% of main-sequence stars at A8. 
 
\begin{marginnote}[]
\entry{Am stars} \  Metallic-lined main-sequence AF stars with overabundances of Fe-peak and rare earth elements, and deficiencies of Ca and Sc.
\entry{magnetic Ap stars} \  Peculiar main-sequence BAF stars with overabundances of rare earth elements up to $10^6$ solar. They have global, roughly dipolar magnetic fields with strengths of $1 - 34$\,kG.
\entry{$\lambda$\,Boo stars} \  Late-B to early-F stars with underabundances of refractory elements.
\end{marginnote}

It was once thought that Am stars do not pulsate as a consequence of He settling in the HeII ionization zone where the $\kappa$-mechanism is the main driver of $\delta$~Sct pulsations \citep{1976ApJS...32..651K}. Later, some Am stars were found to be $\delta$~Sct stars, and a large survey of Am stars with spectra from LAMOST by \citet{2017MNRAS.465.2662S} found that while most Am stars do not pulsate, a fraction of them do in the temperature range $6900 \le T_{\rm eff} \le 7600$\,K at the cooler end of the $\delta$~Sct instability strip. They also found that the stars with the stronger abundance anomalies are less likely to pulsate. \citeauthor{2017MNRAS.465.2662S} proposed that the pulsations in these stars are driven by turbulent pressure \citep{2014ApJ...796..118A}.
 
\citet{2020MNRAS.498.4272M} studied KIC\,11296437, the first known roAp star (Section\,\ref{roAp}) that shows both high-overtone p\,modes typical of roAp stars, and low-overtone p\,modes as in $\delta$~Sct stars. Their models showed how magnetic fields above about 1\,kG damp low-overtone p\,modes, as in $\delta$~Sct stars, and also high overtone g\,modes, as in $\gamma$~Dor stars. Models with He depletion showed that a bump in the Rosseland mean opacity caused by the H-ionization edge allows the $\kappa$-mechanism to drive pulsations in Am stars, with a calculated instability strip that is in excellent agreement with that found observationally by \citet{2017MNRAS.465.2662S}. These models will be tested in the future with asteroseismology of pulsating Am stars. 

\subsubsection{The roAp stars}
\label{roAp}

The rapidly oscillating Ap (roAp) stars are a rare subset of the magnetic Ap stars that pulsate in high overtone magneto-acoustic modes with periods in the range $4.7 - 25.8$\,min. They have strong, global, roughly dipolar magnetic fields with strengths in the range $1 - 34$\,kG. The modes are largely confined to the outer layers of the star. In the visible photosphere and above they are of striking complexity  \citep{2018MNRAS.480.1676Q}. 

I discovered this class \citep{1982MNRAS.200..807K} and later discussed them in this journal \citep{1990ARA&A..28..607K}. 
These were the first stars found to have oblique pulsation axes where the rotational deformation of the star did not dominate the physical selection of the pulsation axis. Many theoretical studies refined and improved the oblique pulsator model that I had proposed. \citet{2002A&A...391..235B} showed that the pulsation axis is not necessarily the magnetic axis, but is still inclined to the rotation axis (see also \citealt{2011A&A...536A..73B}).  \citet{2021FrASS...8...31H} has recently reviewed the roAp stars, and \citet{2021MNRAS.506.1073H} give a detailed introduction to them in their report on TESS cycle 1 observations. I refer the reader to these two papers for the many references to the theoretical and observational literature on roAp stars.

\begin{marginnote}[]
\entry{roAp stars} \  Rapidly oscillating Ap stars pulsate in high-overtone magneto-acoustic modes with periods in the $4.7 - 25.8$\,min range. They are oblique pulsators. 
\end{marginnote}

The importance of oblique pulsation is that the pulsation mode is seen from varying aspect with rotation of the star, providing geometrical information that is not available for stars pulsating along the rotation axis. This has shown that roAp stars pulsate primarily in zonal distorted dipole and quadrupole modes ($\ell = 1,2; m=0$). The magneto-acoustic modes are distorted from pure dipole or quadrupole modes by the combined  Coriolis and Lorentz forces, and the distortion from spherical symmetry that the magnetic field and rotation cause. \citet{2018MNRAS.476..601H} reported four roAp stars with distorted quadrupole pulsations, along with models that include the effect of the magnetic field on the pulsations.  Thus, the roAp stars provide an unusual laboratory to study the interaction of pulsation and magnetic fields on a global scale. The only other place where this physical interaction is observed is in local interactions of p\,modes with sunspots in the Sun.

The Ap stars are the most extreme laboratories where the results of atomic diffusion are observed (Sections\,\ref{diffusion}, \ref{AmandAp}). In the presence of stability against mixing and convection of the outer radiative zone with its strong magnetic field, ions with many absorption lines are radiatively levitated high into the atmosphere to continuum optical depths equivalent to the chromosphere in the Sun, as in the brightest roAp star, $\alpha$\,Cir \citep{2009A&A...499..851K}. In particular, ions of rare earth elements float isolated in cirrus-like clouds at various levels of the outer atmosphere far above optical depth $\tau = 1$, thus resolving the pulsation mode and the atmospheric structure in depth. These ions are also concentrated in spots, governed by the magnetic field configuration, giving surface resolution. Therefore, the roAp stars have the unique potential to map the atmospheric structure and pulsation geometry in 3D, although exploiting this potential is challenging. 

It is possible to resolve the depth dependence of the pulsation mode with radial velocity studies of ions that have line formation layers at different depths, and by studying radial velocity in line bisectors since the spectral line shape maps different depths. Many studies have done this, generally with observations from large telescopes, to meet the combined requirements of high spectral resolution, high signal-to-noise, and high time resolution. This work was pioneered by \citet{1998MNRAS.300L..39B}, \citet{2001A&A...374..615K}, and \citet{2003MNRAS.345..781M}.

\citet{2006A&A...446.1051K} studied the atmospheric depth behavior of pulsation in the photometrically well-studied, singly periodic roAp star HR\,3831 with a spectroscopic study of line profile variations. He noted the superior diagnostic value of studies using radial velocity variation in spectral lines formed at different atmospheric depths compared to the blunter tool of photometric observations. In particular, he warned that the use of the amplitudes of frequency multiplets generated by rotation can only be used in the oblique pulsator model to infer a geometry of the pulsation mode at the depth of the line forming region for a particular line, and at the surface position of the spots, for ions confined to those. These differ for different lines.

This problem has become clear in photometry from TESS observations of roAp stars, particularly HD\,6532 \citep{2020ASSP...57..313K}, and also other stars \citep{2021MNRAS.506.1073H}. The amplitudes of the multiplet frequencies generated by oblique pulsation provide information about the mode geometry. However, these are very different for observations made through the TESS broad-band filter ($0.6 - 1.0$\,$\upmu$m) compared to ground-based observations of the same star typically through a Johnson $B$ filter (effective central wavelength $0.4353$\,$\upmu$m), much bluer than the TESS filter. Since the continuum opacity of the stellar atmosphere is wavelength dependent, observations through different filters sample different atmospheric depths. Interpretation of the geometry of the pulsation mode thus requires the same caution pointed out by \citet{2006A&A...446.1051K} from higher geometrical resolution spectroscopic observations of HR\,3831. 

{Section\,A.8 in the appendix} gives two examples of high resolution studies of roAp stars, discusses discoveries from TESS for roAp stars, and gives further details of the complexities of asteroseismology for these stars. Those high resolution spectroscopic studies stimulated the work of \citet{2018MNRAS.480.1676Q}, who developed a theoretical model of roAp stars with coupling between the acoustic pulsations and the magnetic field. They showed in detail how the pulsation amplitude increases with atmospheric height and how the pulsation phase changes with height and viewing angle. The roAp stars are difficult. Their atmospheric abundances are stratified, the temperature gradient is not normal, the magnetic field concentrates some elements in surface spots that are difficult to model uniquely,  and the pulsations are magneto-acoustic with strong changes in pulsation amplitude and phase in all dimensions. Understanding these complexities with further theory and full rotational high resolution spectroscopic observations promises the most detailed view of pulsation and atmospheric abundance distributions for any stars other than the Sun. 

\subsubsection{$\lambda$\,Boo stars}
\label{lambdaboo}

 The $\lambda$\,Boo stars are generally rare, constituting only about 2\% of late-B to early-F stars. Strikingly, however, they constitute more than 30\% of Herbig Ae pre-main-sequence stars (Section\,\ref{pms}), which host proto-planetary disks.  The $\lambda$\,Boo stars are not slow rotators, having a mean $v \sin i \sim 160$\,km~s$^{-1}$, similar to that for spectroscopically normal A stars (i.e., stars that are not Am, Ap, or $\lambda$\,Boo). They show deficiencies down to $10^{-2}$ in the abundances of refractory elements (those with high melting points), such as Fe, Mg, and Si, while volatile elements (with low melting points), such as C, N, and O, have normal abundances. The definition of the class of $\lambda$\,Boo stars has been in flux, which \citet{2015PASA...32...36M} have clarified. 

Hypotheses for the $\lambda$\,Boo phenomenon include accretion of interstellar or circumstellar gas.  \citet{2015A&A...582L..10K} presented a model where giant planets sweep up dust so that the star accretes gas that is depleted in refractory elements. They showed a correlation between $\lambda$\,Boo abundances in Herbig Ae stars and the presence of gaps in their proto-planetary disks, and suggested that diversity in planet formation leads to the range of abundances observed in $\lambda$\,Boo stars. 

About 80\% of $\lambda$\,Boo stars are also $\delta$~Sct pulsators, hence can provide asteroseismic information, in particular age, mass, and metallicity. An example is HD\,139614, a $\lambda$\,Boo/$\delta$~Sct star that has a proto-planetary disk with gaps. It is in the Upper Centaurus--Lupus part of the Scorpius--Centaurus association, the nearest region to the Sun, at 140\,pc, where there is massive star formation. There is a range of ages for the stars across the association, with a median age of 16\,Myr and a 1$\sigma$ spread of 7\,Myr. \citet{2021MNRAS.502.1633M} found an asteroseismic age for HD\,139614 of $10.8 \pm 0.8$\,Myr. This is the most precise age determination for a star in the association, and is in good agreement with other age determinations, showing the power of asteroseismology for these pre-main-sequence stars. \citeauthor{2021MNRAS.502.1633M} also found a normal metallicity for their best model of HD\,139614, which is the first determination that $\lambda$\,Boo stars have globally normal metallicity, as is expected from the accretion model for their surface abnormalities. 

\citet{2020Natur.581..147B} found from asteroseismology and space motions that some of the 60 $\delta$~Sct stars for which they determined the large separation, $\Delta \nu$, and which they modeled, are young (Section\,\ref{deltasct}). They found an occurrence rate of 10\% for $\lambda$\,Boo stars in their sample of 60 stars, supporting the hypothesis that many $\lambda$\,Boo stars are young and accrete from dust-depleted circumstellar disk material.  Nevertheless, in a survey of southern hemisphere $\lambda$\,Boo stars, \citet{2020MNRAS.499.2701M} found only about 40\% of them are young, with the rest further evolved along the main-sequence. If the disk accretion model is correct, then the surface convection zone must not mix away the accreted chemical abnormalities over the main-sequence lifetimes. More modeling is needed to determine if sufficient dust-depleted gas can accrete to provide the abundance anomalies through the entire surface convection zone, which has a thickness of only thousands of kilometers. Clearly, the origin of the $\lambda$\,Boo phenomenon still has its challenges. Asteroseismic ages and global metallicities for these stars will contribute to deeper understanding of them.

\subsubsection{Summary: Pulsation in chemically peculiar stars}
\begin{enumerate}
\item Pulsation in metallic-lined A (Am) stars is common with relatively low amplitudes. Stars with stronger abundance anomalies are less likely to pulsate due to depletion of HeII driving.
\item Pulsation in rapidly oscillating Ap (roAp) stars is rare, even at the $\upmu$mag detection level. 
\item The oblique pulsator model shows the mode geometries in roAp stars to be complex as a function of atmospheric height. Spectroscopic line-by-line studies provide 3D information on atmospheric structure. 
\item Young $\lambda$\,Boo stars appear to be accreting planetary disk material. They are usually $\delta$~Sct stars for which asteroseismic ages are in agreement with gyrochronology.
\end{enumerate}

\subsection{\ Pre-main-sequence pulsators}
\label{pms}

Historically, pre-main-sequence evolution was thought of in terms of a star of a given mass having an evolution track across the HR~Diagram, first down a convective Hayashi track, then across towards the main-sequence as the envelope becomes more radiative, until H fusion is ignited and the zero-age main-sequence is reached. It has been clear for decades that pre-main-sequence evolution is much more complex. Stars form as seeds, or cores, in accretion envelopes. During their contraction phases they accrete mass, so that no single mass can be attributed to a star along its pre-main-sequence pathway in the HR~Diagram. See, e.g., \citet{1991ApJ...375..288P}, \citet{2004fost.book.....S}, \citet{2017A&A...599A..49K}.

\begin{marginnote}[]
\entry{Herbig Ae/Be stars} \  Early-B to early-F pre-main-sequence stars showing emission lines. Ages $< 10$\,Myr. Some are $\gamma$~Dor, $\delta$~Sct, or SPB stars.
\end{marginnote}

As these pre-main-sequence stars emerge from their cocoons at the stellar birth-line, and appear observationally on the HR~Diagram, they still have circumstellar material, particularly an accretion disk. High mass stars that are born as early-B stars do not emerge from their cocoons until about the time they ignite H fusion. Low mass, $M \le 1$\,M$_\odot$, pre-main-sequence TT\,Tau stars may have stochastically excited modes, but those frequencies are generally lost in the FT because of the variations from the circumstellar material. Recently, attempts to extract the solar-like oscillation signal from the accretion noise have been made \citep{2021A&A...647A.168M} with some possible candidates found. 

It is the stars between these high and low masses for which there is great promise for asteroseismology of pre-main-sequence stars. These are Herbig Ae and Be stars \citep{1960ApJS....4..337H} with masses $1.5 \le M \le 4$\,M$_\odot$.  As they contract, some of them appear as $\delta$~Sct, $\gamma$~Dor, and SPB pulsators prior to their arrival on the zero-age main-sequence.  The opportunity to observe the interior of these stars asteroseismically to test and refine pre-main-sequence models is attractive and promising.

A problem in fulfilling that promise is that variable dust in the line of sight causes irregular light variations from which it can be difficult to extract pulsation signals. This is particularly true for $\gamma$~Dor stars which have low frequencies in the range where the noise from the disk variations is strong. The p\,mode frequencies, however, can be separated in the FT from the disk noise, as first shown by \citet{1995MNRAS.276..191K} for HR\,5999 and in more detail for HD\,142666 by \citet{2009A&A...494.1031Z} with photometry from the MOST mission. While \citet{2008ApJ...673.1088Z} listed only 36 pre-main-sequence pulsators and candidates, that number has now grown significantly. The goal with these stars is to distinguish asteroseismically their internal structure from the zero-age main-sequence structure stars such as they will soon become. 

Some pre-main-sequence $\delta$~Sct stars have planet forming disks that do not cause irregular light variations, hence have cleaner FTs at low frequency. An example is HD\,139614, a $\lambda$\,Boo star (\citealt{2021MNRAS.502.1633M}; Section\,\ref{lambdaboo}). \citet{2021A&A...654A..36S} have found in TESS data 16 new pre-main-sequence $\gamma$~Dor, $\delta$~Sct, and SPB stars (or candidates) and initiated a modeling program for all known pre-main-sequence stars. They provide a good introduction to the current state of asteroseismology of these stars.  See also \citet{2020svos.conf...39Z} for a short review of this field. 

\section{The upper classical instability strip: RR\,Lyr stars and Cepheids}
\label{classical_is}

The classical pulsators of the upper instability strip are fundamental to the galactic and extra-galactic distance scales. These are, broadly, the RR\,Lyr stars and the Cepheid variables. Both classes have an array of subtypes, some of which are noted in Figure\,1. The Cepheids are divided into Type\,I (population\,I) and Type\,II (Population\,II) Cepheids; Anomalous Cepheids or BL\,Boo stars; BL\,Her and W\,Vir stars (also Population\,II); and RV\,Tau stars (also Population\,II).  These types are based on light curves, on the Period-Luminosity relation, and on the mass, age and evolutionary stage of the stars. The RR\,Lyr stars also have subtypes: RRab, RRc, RRd, and stars showing the Tseraskaya--Blazhko effect. See \citet{2010aste.book.....A} for a guide the the nomenclature and evolutionary stages of the subtypes. 

\begin{textbox}[h!]\section{The Tseraskaya-Blazhko effect}
The longest standing puzzle for RR\,Lyr stars concerns amplitude and period changes that have been called for over 60 years the Blazhko effect, based on an observation of RW Dra \citep[RW\,Dra = Var. 87.1906 Draconis]{1907AN....175..325B}. Sergey Blazhko worked at the Moscow Observatory, where he was later director. I note  in the opening sentence of this highly-cited paper that Blazhko says that the observational changes in period were discovered by ``Mrs Ceraski''. Lidiya Petrovna Tseraskaya ($1855 - 1931$) worked at Moscow Observatory (now the Sternberg State Astronomical Institute), where she discovered 219 variable stars. It was she who published under the name Mrs W. Ceraski, taken from a transliteration from Cyrillic of her husband's form of their surname. Today, Mrs W. Ceraski would be included as co-author Lidiya Tseraskaya, as it was she who made the discovery. ADS lists 155 observing notes on variable stars published by her between $1879 - 1914$. In 1908 she was awarded the prize of the Russian Astronomical Society, and crater Tseraskaya on Venus is named after her. I propose that what has previously been known as the Blazhko effect, based on this one paper, would more appropriately be known as the Tseraskaya-Blazhko effect.
\end{textbox}

The Period-Luminosity relation, or Leavitt Law, for Cepheids was first noted by Henrietta Leavitt in 1908 and then announced with confidence in 1912 \citep{1912HarCi.173....1L} when she commented, ``A remarkable relation between the brightness of these variables and the length of their periods will be noticed ... the brighter variables have the longer periods.'' The physical origin of this relation is now known from asteroseismology: the brighter stars have  longer sound travel times across the star, hence longer pulsation periods, while the fainter stars have shorter sound travel times, hence shower periods, both as a consequence of radius and density.  Sound travels more slowly and has farther to go in the brighter Cepheids. It is this Leavitt Law that makes the Cepheids so useful as distance indicators, and it is primarily used on fundamental radial mode, or first overtone mode, pulsation. 

But Cepheids and RR\,Lyr stars are much more interesting asteroseismically than simple low-overtone radial mode pulsators. They are high amplitude pulsators with strongly nonlinear light curves that challenge and test nonlinear pulsation theory. A significant fraction of them pulsate in two modes, or more, some of which are nonradial, providing the data for asteroseismic modeling. They show stable frequencies in some stars, frequency jitter in others, period doubling (alternating maxima in the light curves), the presence of subharmonics, and alternating minima in late evolutionary stage RV\,Tau stars (see, e.g., \citealt{2021ApJS..253...11P}). Most, or all of these dynamical effects are related to mode resonances, providing additional asteroseismic constraints on the stellar models. 

\begin{marginnote}[]
\entry{Tseraskaya-Blazhko effect} \  The modulation of the light curves of RR\,Lyr stars on times scales of 10s to 100s of days.
\end{marginnote}

A large fraction of RR\,Lyr stars show the Tseraskaya--Blazhko effect -- clear modulation of the light curves on times scales of 10s to 100s of days. The modulation is not simply periodic. \citet{2017MNRAS.466.2602P}, in a massive survey using OGLE-IV data, examined almost 8300 RR\,Lyr stars in the Galactic Bulge, finding that over 40\% of them show Tseraskaya--Blazhko modulation of their light curves. While the K2 light curves for RR\,Lyr stars are not fully analyzed, they give a similar incidence. The Tseraskaya--Blazhko effect is thought to be the result of a resonance, and there is some connection to period doubling in some stars, supporting that idea. It is remarkable that after more than a century and with all of our powerful asteroseismic modeling tools, this phenomenon is not fully understood in stars of such fundamental importance.

RR\,Lyr stars have periods of about $0.5 -1.2$\,d and Cepheids $1 - 100$\,d, making it possible to study these stars with ground-based survey projects, e.g., OGLE-IV \citep{2015AcA....65....1U}, among others. They will soon benefit further with upcoming data from the Zwicky Transient Facility (ZTF, see {Section\,B.2 in the appendix}) and Vera Rubin Observatory (VRO) surveys. Recent literature on these classical variables is vast and growing, with newly observed behavior, and new theoretical interest. While the ground-based surveys dominate the data, with light curves for thousands of stars, space mission data from MOST, CoRoT, {\it Kepler}, TESS, and BRITE have illuminated our view of pulsation in these upper instability strip stars, as a consequence of long, uninterrupted high-precision light curves. 

An outstanding tool for organizing the great variety of pulsation frequency behavior in Cepheids and RR\,Lyr stars is the Petersen Diagram \citep{1973A&A....27...89P}, which plots the period ratio of two frequencies against the frequency of the longer period mode. This diagram distinguishes, e.g., fundamental to first overtone radial mode pulsators from higher overtone pulsators;  nonradial pulsators; stars with period doubling; and a variety of as-yet not understood pulsation behavior.  \citet{2017EPJWC.15206003S} discussed the period-doubling phenomenon, first seen in {\it Kepler} data of RR\,Lyrae itself \citep{2010ApJ...713L.198K}. This phenomenon leads to alternating maxima in the light curves as a result of a frequency resonance, asteroseismically constraining models of these stars. \citet{2017EPJWC.15206003S} give an excellent short review entitled the ``Petersen Diagram Revolution''.  \citet{2021FrASS...7...81P} reviewed {\it Kepler} observations of RR\,Lyr stars. Their figure\,8 further illuminates the wide variety of pulsation behavior seen in the Petersen diagram for these stars. 

\section{After the red giants: asteroseismology of white dwarfs, subdwarf O and B stars, and extreme He stars}

\subsection{Pulsating white dwarfs}
\label{whitedwarfs}

Many white dwarfs carry information about the cores of their progenitor red giant stages, including chemical changes from nuclear reactions and angular momentum transfer during early evolutionary stages. Some of them have a modicum of nuclear energy generation via hydrogen shell-burning, but on the whole they radiate left-over energy. That comes in several fascinating forms. Thermal energy remaining from the hot red giant core in the non-degenerate nuclei is a major source of white dwarf luminosity, but that is not always dominant. White dwarf cores are  typically C-O, but also O-Ne among the ultra-massive white dwarfs with $M \ge 1.05$\,M$_\odot$. When white dwarfs cool sufficiently, the cores crystallize leading to a release of latent heat. As a result of Coulomb interactions, the nuclei are forced into a crystalline structure, unlike Earthly crystals, which are bound by electron sharing; unbound electrons cannot be shared in a degenerate gas. Some white dwarfs are now known to have cores that are more than 90\% crystallized \citep{2019A&A...632A.119C}, an amazing thought: Single crystals roughly the size of the Earth. Asteroseismic results support this \citep[e.g.,][]{2019A&A...621A.100D}. 

In some white dwarfs the neutrino luminosity is the dominant radiation loss. If the neutrino has a magnetic moment -- an important question for the Standard Model of particle physics -- then an abundant source of neutrinos in white dwarf stars comes from plasmon decay of photons to neutrino-antineutrino pairs as a consequence of neutrino-photon interaction. White dwarf cooling times provide a direct measure of this process, and asteroseismology measures those cooling times.  If axions exist, providing an explanation for the charge-parity problem in strong interactions, it is possible that white dwarf cooling times may also constrain axion mass, a quantity not predicted by theory. Asteroseismology measures white dwarf cooling times by determining evolutionary period changes to the pulsations (although, see my caution about evolutionary weather in Section\,\ref{freqvar}). See \citet{2008PASP..120.1043F}, \citet{2008ARA&A..46..157W}, and \citet{2019A&ARv..27....7C} for extensive reviews of white dwarfs, including pulsations.

\citet[chapter 2]{2010aste.book.....A} discussed in detail the arcane naming conventions for white dwarf stars. Briefly here: in order of decreasing temperature, the DO, DB, and DA white dwarfs have spectral features similar to those of O, B, and A main-sequence stars, but not necessarily the same effective temperatures. Pulsating variables of those three spectral classes are the DOV (GW\,Vir), DBV (V777\,Her), and DAV (ZZ\,Cet) stars that are the primary asteroseismic classes. There are also three additional classes of variable white dwarfs that are of asteroseismic interest: the DQV, hot DAV, and Ultra-massive DAV stars. 

Stars in close binary systems interact. In some cases a companion strips away the outer envelope of a star leaving behind a type of white dwarf that could never form as a consequence of single star evolution. Extremely low-mass (ELM) white dwarfs with He cores are exciting examples of this, and there are classes of ELM white dwarfs that pulsate and can be studied asteroseismically. Now, mostly in the last decade, we add to the DOV, DBV, and DAV classes the extremely low-mass variables (ELMV) and pre-ELMV, and some other rare (sometimes contested) classes. In other binary cases material from a companion is accreted by a white dwarf, giving a born-again blue straggler. And, of course,  almost the entire class of cataclysmic variables is made up of close interacting binary stars with one star a white dwarf. Many of these are the progenitors of novae and dwarf novae. Some of those also pulsate and may be amenable to asteroseismic inference, e.g., GW\,Lib \citep{2004MNRAS.350..307V, 2021MNRAS.502..581C}. Finally, in some binary pairs with two white dwarfs, orbital energy losses lead to mergers and the production of exotic stages in the lives of a few stars, such as the eHe stars (Section\,\ref{pvtel}), a few of which pulsate spectacularly.

The {\it Kepler} main mission did not observe many white dwarf stars, as it was targeting a relatively distant part of the Galaxy just out of the plane and white dwarfs in that direction are too faint. However, the extended {\it Kepler} mission, K2, did target white dwarf stars; see \citet{2020FrASS...7...47C} for a specific review of the {\it Kepler} results for asteroseismology of white dwarfs. And now TESS is also observing white dwarfs. While both {\it Kepler} and TESS missions are known for the high precision of their photometry, that is not greatly better than ground-based photometry for white dwarfs because the faintness of these stars and the small (by ground-based standards) apertures of the space telescopes mean that the noise is photon limited. Additionally, both the {\it Kepler} and TESS broad-band filters do not incorporate blue to UV observations where most white dwarf stars have maximum pulsation amplitude. The advantages of the {\it Kepler} and TESS data sets come from their long time spans and continuity. Gaia has now detected hundreds of thousands of new white dwarfs, and white dwarf candidates, some of which will be studied by TESS, and others, later in this decade, by the ESO space mission PLATO\footnote{PLATO: https://sci.esa.int/web/plato}. 
 
White dwarfs are fascinating in themselves, and they have importance beyond stellar astrophysics. White dwarfs with a mass near the Chandrasekhar limit\footnote{See {Section\,B.1 in the appendix} for a history of the Chandrasekhar -- Eddington debate.} that orbit a donor companion in a close binary star are the progenitors of the Type Ia Supernovae that are the standard candles of cosmology, fundamental to models with acceleration of the expansion rate and a need for Dark Energy. Single white dwarfs are excellent clocks to measure the age of galactic disk components and star clusters, given a knowledge of their cooling times, which asteroseismology can constrain. Hence pulsating white dwarfs are of broad interest. See {Sections\,B.2 -- B.6, in the appendix} for detailed discussions of asteroseismology of white dwarf classes DAV (ZZ\,Cet), DOV (GW\,Vir), DBV (V777\,Her), ELMV (Extremely Low Mass Variables), and BLAPs (Blue Large Amplitude Pulsators), respectively. See {Section\,B.7 in the appendix} for discussion of the DQV, hot DAV, and Ultra-massive DAV stars.

\subsection*{Summary: Asteroseismology of white dwarfs}
\begin{enumerate}
\item  Asteroseismology of white dwarfs provides measurements of mass, radius, temperature, luminosity, age, rotation, and the extent of the non-degenerate surface layers, and tests the physics of matter under extreme conditions.
\item White dwarfs are progenitors of type Ia supernovae, the standard candles of observational cosmology.
\item Neutrinos are the dominant form of radiation from some white dwarfs, providing observational tests of neutrino physics. 
\item Extremely low mass white dwarfs are the stripped cores of red giants, allowing direct observations of red giant core physics. 
\item White dwarfs in binaries are important in the study of cataclysmic variables, SN\,Ia precursors, blue stragglers, and exotic results of mergers, such as extreme He stars.
\end{enumerate}

\subsection{Extreme Horizontal Branch stars: The subdwarf O and B variables}
\label{sdbv}

There is a population of He core burning stars with thin H envelopes on the extreme, or blue horizontal branch with typical masses around 0.5\,M$_\odot$ -- the subdwarf B (sdB) stars. About half of them are in compact binaries; they are the stripped cores of red giants after common envelope ejection, or Roche lobe overflow. The single sdB stars probably originate from He white dwarf mergers, but possibly from other channels, including white dwarf -- low mass main-sequence star mergers, and rarely mergers involving neutron stars \citep{2020A&A...634A.126W}. Some of them are also progenitors of the ELM white dwarfs (see {Section\,B.5 in the appendix}).  Most evolve to subdwarf O (sdO) stars when helium is exhausted in their cores and they begin He shell burning, a stage with a lifetime much shorter than the $\sim$$10^8$\,yr of He core burning, hence sdO stars are rarer than sdB stars. 

The evolutionary relation of sdO and sdB stars is not completely clear. Most of these stars continue to evolve directly to the white dwarf stage without a return to the Asymptotic Giant Branch. The exceptions are those with white dwarf companions that become Type Ia supernovae. The thin H atmospheres ($10^{-2}$\,M$_\odot$) of the sdB stars are He deficient as a consequence of gravitational settling (section\,\ref{diffusion}).  The sdO stars, on the other hand, sometimes show He deficiency, sometimes not. These subdwarf stars are fascinating in their own right, and important for our understanding of red giant and white dwarf evolution. See \citet{2016PASP..128h2001H} for a comprehensive introduction to, and review of these stars. See {Section\,C in the appendix} for a history of the discovery of the sdB stars and a detailed discussion of the asteroseismology of sdB and sdO stars.

\subsection*{Summary: sdBV stars}
\begin{enumerate}
\item Asteroseismology of sdBV stars has provided the fundamental parameters mass and radius, as well as structural parameters -- the envelope mass, the core mass and the core composition -- illuminating their evolutionary origins. 
\item Asteroseismic measures of the core O mass fraction in sdBV stars may ultimately constrain the $^{12}{\rm C}(\alpha,\gamma)^{16}{\rm O}$ reaction rate in these stars.
\end{enumerate}

\subsection{Extreme Helium (EHe) stars,  PV\,Tel stars, and R\,CrB stars}
\label{pvtel}

Not all pulsating stars are amenable to asteroseismology. Irregular variables provide only some astrophysical inference. Singly periodic variables -- even when mode identification is certain -- provide only a single constraint for forward modeling. Nevertheless, some of these stars are in such exotic evolutionary stages that they deserve mention in this review. These are the extreme He (EHe) stars, which have unusually high luminosity-to-mass ratios, little or no H, and enhanced C in their He atmospheres.

The evolutionary pathway to EHe stars is that of a binary merger of a typical 0.6-M$_\odot$ C-O white dwarf with an 0.3-M$_\odot$ low mass He white dwarf. Ignition of a He shell flash results in a supergiant with strong mass loss and a He-burning shell that eats its way inwards in the core of the star \citep{2002MNRAS.333..121S}, remarkably in the opposite direction to normal shell burning. This pathway produces R\,CBr stars \citep{2019ApJ...885...27S}, with their episodes of dramatic drops and recoveries in optical light when C soot forms in their atmospheres and then is ejected by radiation pressure. Evolution of EHe stars is then  blueward, eventually to become a pre-white dwarfs, then white dwarfs. These rare stars are inherently interesting for their novelty.

\begin{marginnote}[]
\entry{EHe stars} \ high $L/M$ He stars evolved from merged white dwarfs.
\entry{PV\,Tel stars} \ Pulsating EHe stars.
\entry{The born-again rocket star} \ V652\,Her accelerates pulsationally from $0 - 100$\,km~s$^{-1}$ in 15\,min with a supersonic Mach-10 pulse.
\end{marginnote}

Among the EHe stars are variables known as PV\,Tel stars with generally irregular light curves with timescales of $0.5 - 25$\,d.  The class is disparate, hence, despite small numbers, \citet{2008IBVS.5817....1J} divided it into three subtypes, the third  (his type III) of which are the BX\,Cir stars comprising only two stars, V652\,Her and BX\,Cir, both of which have regular pulsations with periods of 2.4\,hr. 

V652\,Her is the better studied star, hence is labelled in Figure\,1, which shows that it lies somewhat below the main-sequence. Its evolutionary future is to evolve through a series of loops to become a pre-white dwarf. At present, it is the remarkable ``born-again rocket star'' (\citealt{2015MNRAS.447.2836J}; \citealt{2016A&G....57d4.37K}) which accelerates pulsationally from $0 - 100$\,km~s$^{-1}$ in 15\,min with a Mach-10 pulse (239 km~s$^{-1}$) running through its atmosphere! This, thus, allows a look at extreme pulsation amplitude and supersonic velocity in a stellar atmosphere. At present asteroseismology does not particularly exploit pulsation amplitude, as the nonlinear theory needed to understand fully what determines pulsation amplitude is difficult. (An exception to that is equipartition of driving energy in stochastic oscillators which allows inference of viewing angle from relative amplitudes of the components of dipole multiplets.) In looking to a future when driving and damping may be modeled to explain successfully what determines mode amplitudes, the BX\,Cir stars will provide the extreme limits of pulsation amplitude to test the theory. While there are only two such stars currently known, more should be found in TESS data. 

\citet{2020MNRAS.495L.135J} studied two EHe stars observed with TESS, both of them irregular variables. They concluded that the variability time scale is of the same order as the radial fundamental mode. The reason for the irregularity is that the pulsations are non-linear, non-adiabatic, and possibly not spherically symmetric -- an interesting challenge!

\section{Tides: Asteroseismology of close binary stars}
\label{closebinaries}

Asteroseismology is simplest for stars that are approximately spherically symmetric. For slowly rotating single stars with a weak, or no magnetic field, one-dimensional models are adequate. In recent years two-dimensional models of stellar pulsation have been developed for the case of rapidly rotating stars \citep{2021A&A...645A..46R}, with potential application to the more rapidly rotating upper main-sequence $\delta$~Sct, SPB, and $\beta$\,Cep stars. Pulsating stars in wide binaries are asteroseismically modeled as single stars. Where they are in eclipsing binaries the determination of fundamental parameters -- particularly mass and radius -- shows that model-dependent asteroseismic determinations of those quantities have excellent accuracies, attaining a few percent in the best cases. This gives confidence in other asteroseismically determined quantities, particularly age and global metallicity.

Upper main-sequence stars mostly exist in binary or multiple star systems, with the incidence increasing with increasing mass to more than 90\% for the hottest O stars \citep{2017ApJS..230...15M}. If we include the high incidence of planetary systems deduced from results of {\it Kepler} and TESS, then single stars with no stellar or planetary companion are rare. The stars of interest in this section are those that orbit closely enough that tidal forces cannot be neglected, for which the asteroseismic problem becomes more challenging, and offers new inferences on the physics of close binary stars. 

Using pulsating stars as stable frequency standards  has led to new ways to study binary star orbits: the frequency modulation (FM; \citealt{2012MNRAS.422..738S}, \citealt{2015MNRAS.450.3999S}) and phase modulation (PM; \citealt{2014MNRAS.441.2515M},  \citealt{2016MNRAS.461.4215M}) techniques. In addition, the orbital motion of the {\it Kepler} satellite allows the Nyquist limitation in frequency analysis of uniformly spaced data to be overcome in the super-Nyquist asteroseismology technique (sNa;  \citealt{2013MNRAS.430.2986M}). See {Section\,D in the appendix} for a discussion of these techniques and their impact. 

\citet{2021FrASS...8...67G} has given an extensive review of asteroseismology in close binary stars with particular emphasis on tides and the effects of mass transfer on pulsations and stellar structure. He examined both tidally excited and tidally perturbed, self-excited oscillations. \citet{2021Galax...9...28L} has  also reviewed eclipsing binaries, both wide and semi-detached, with pulsating components, giving individual examples of $\delta$~Sct, $\gamma$~Dor, and $\beta$\,Cep stars observed by {\it Kepler} and TESS.  The reader is referred to these reviews for thorough discussions of this exciting new field of asteroseismic research.

Close binaries lead to: mass loss; mass interchange; common envelope evolution; spin-orbit synchronization; orbital and rotational energy exchange with pulsation; and, of course, tides. They are the progenitors of type Ia supernovae, the distance indicators of choice in the deep universe; of binary neutron stars and black holes, the sources of measurable gravitation radiation, gamma ray bursts, and many of the heavier elements; of cataclysmic variables; of blue stragglers; of X-ray binaries; of the newly discovered heartbeat stars, and the tidally tilted pulsators; and of Doppler beaming and boosting, which have only become widely observable with the high precision of the space photometry. 

Doppler beaming and boosting has multiple components. The simple one is boosting where a component of a binary star appears brighter when approaching and fainter when receding simply because of the Doppler shift altering the energy in the photons. Add to that the special relativistic effects of beaming, time dilation, and photon arrival time and we expect all orbiting stars to be brighter when approaching and fainter when receding than they would otherwise have been without these effects. Prior to the precision of space photometry, Doppler beaming was difficult, or impossible, to observe as its amplitudes are typically only hundreds of $\upmu$mag. A beautiful demonstration of it was given in the study of {\it Kepler} data of the eclipsing binary KPD\,1946+4340, which consists of an sdB star and a white dwarf in a 9.7-hr orbit \citep{2011MNRAS.410.1787B}. Clearly visible in the light curve are ellipsoidal variation from tidal distortion, the Doppler beaming modification of the ellipsoidal variability, the eclipses, and even the general relativistic effect of gravitational lensing by the white dwarf during primary eclipse. For space photometry of eclipsing binaries, Doppler beaming is now a standard part of light curve modeling. This is a case (see {Section\,D.1 in the appendix} for another) where it is now possible to measure radial velocities without the need for spectroscopy using the Doppler beaming signal. This saves expensive telescope time, and even enables studies where rotational broadening of spectral lines and/or faintness of the target preclude obtaining radial velocities spectroscopically. 

Systematic searches for eclipsing binaries with pulsating components are ongoing. \citet{2019A&A...630A.106G} reported 303 pulsating eclipsing binaries from a systematic search of nearly 3000 eclipsing binaries observed by {\it Kepler} \citep{2016AJ....151...68K}. \citet{2021A&A...652A.120I} reported over 3000 eclipsing binaries and candidates from an automated search of about 190\,000 upper main-sequence stars in nearly all-sky data from TESS. EL\,CVn binaries have a bloated pre-He white dwarf in a short-period orbit about an AF star that in some cases is a $\delta$~Sct star. More than 100 of these systems are now known, in two of which the white dwarf pulsates. See {Section\,B.8 in the appendix} for further discussion of these fascinating binary systems. An explosion of discovery from pulsating  binary stars is imminent. 

\subsection{Heartbeat stars and tidal locking}
\label{hbstars}

\citet{1995ApJ...449..294K} presented a theoretical study of tidal excitation of pulsation modes in eccentric binary systems, applying their results to two binary pulsars, although the relative faintnesses of those systems made it difficult to test the theory conclusively. They showed that the tidal force should excite pulsations with mode frequencies that are exact harmonics of the orbital frequency. There were observational indications of this behavior found in ground-based studies (e.g., \citealt{2002MNRAS.333..262H}), but the breakthrough came with the discovery of the first heartbeat star, HD\,187091, better known as Kepler Object of Interest (KOI) 54 \citep{2011ApJS..197....4W}. This star shows sharp increases in brightness with a period of 41.8\,d with commensurate pulsations superposed. The two highest amplitude pulsations are g\,modes with frequencies that are exactly harmonics 90 and 91 of the orbital frequency. It was the continuity and high precision of the space data that made the tidally induced behavior obvious in the light curve. 

\begin{marginnote}[]
\entry{Heartbeat stars} \ Highly eccentric binary stars showing a regular pulse in brightness during periastron passage. More than 20\% pulsate in tidally excited modes. 
\end{marginnote}

Following the discovery of KOI-54, \citet{2012ApJ...753...86T} announced the class of heartbeat stars in a study of 17 of these eccentric binaries in {\it Kepler} data. There are now 172 known in the {\it Kepler} Eclipsing Binary Catalog \citep[KEBC;][]{2016AJ....151...68K}, and more than 200 have been discovered in TESS data. The regular brightening in the light curves of these stars is reminiscent of an electrocardiogram, hence the heartbeat name; more prosaically, they are just highly eccentric binaries. For non-eclipsing systems, the orbital variations are dominated primarily by the tidal distortion, with contributions from heating by reflection and Doppler beaming. The orbital periods are usually under 100\,d and eccentricities range from about 0.2 to 0.9! In the highest eccentricity systems the periastron passage is within a few stellar radii, with stunning tidal distortions that then relax as the stars separate in the outer parts of the orbit. See \citet{2021FrASS...8...67G} and \citet{2020ApJ...903..122C} for example light curves, FTs, and detailed discussion for some of these stars. Two results of the discovery of the heartbeat stars are that the previously untested theory of \citet{1995ApJ...449..294K} provides excellent fits to the wide variety of the orbital light variations that result primarily from different viewing angles and eccentricity, and that strong astrophysical inference about tidally excited oscillations can be made in the more than $20$\% of these systems that show pulsations. 

The tidal resonance between pulsation and orbital harmonics can be very finely tuned and of astrophysical importance. \citet{2018MNRAS.473.5165H} showed that the heartbeat star KIC\,8164262 has an orbital period of $P_{\rm orb} = 87.5$\,d,  an eccentricity of $e = 0.87$, and a prominent pulsation at exactly 229 times the orbital frequency. While there are other oscillation modes, this one very high harmonic resonance dominates the pulsation in the light curve. In a companion paper, \citet{2017MNRAS.472L..25F} discussed the theory for this particular heartbeat star, while \citet{2017MNRAS.472.1538F} gave a general theoretical examination of the physics of excitation and locking caused by dynamical tides. The heartbeat stars support the thesis that there is a positive feedback between the orbital evolution and stellar evolution that keeps the pulsation frequency resonant with a high harmonic of the orbital frequency. This is tidal locking. As the star and orbit evolve, the pulsation cavity and rotation of the star change, modifying the natural mode frequencies, while tidal dissipation causes orbital decay and an increase in the orbital frequency. Feedback keeps the pulsation frequency locked to an orbital harmonic, giving a stable, high amplitude tidally excited oscillation, as in KIC\,8164262. 

While tidal locking does not require high eccentricity -- indeed, for circular orbits, ellipsoidal variability arises from a co-rotating quadrupolar tidal distortion -- it is most easily seen in some pulsating heartbeat stars. In a study of tidal excitation in heartbeat stars \citet{2020ApJ...903..122C} found only 2 of 4 stars studied to exhibit tidal locking; i.e., not all do. The study of tidal locking in heartbeat stars informs other occurrences, such as migration in planetary and moon systems, and inspiralling in white dwarfs. \citet{2013MNRAS.433..332B} examined double degenerate white dwarf binaries where gravitational radiation results in orbital decay. As in some heartbeat stars, they found resonance locking between the orbital frequency, spin frequency, and tidally excited g\,modes. This important result is much more difficult to test observationally in faint white dwarfs, but these are interesting connections. 

Pulsating heartbeat stars are most commonly found among AF stars because of selection effects, although \citet{2014A&A...564A..36B} found and studied 18 heartbeat red giants with eccentricities $0.2 - 0.76$, and with orbital periods up to 440\,d. Pulsation in heartbeat stars is difficult to detect for stars less massive than about 1.5\,M$_\odot$ since the g\,modes are not visible through the thick surface convection zone -- a well-known problem in the Sun. Even though g\,modes are common in $\gamma$~Dor, $\delta$~Sct, and SPB stars, the initial mass function and shorter lifetimes mean that the more massive stars are more uncommon. \citet{2021A&A...647A..12K} made a search of TESS sectors $1 - 16$ for heartbeat stars among $M \ge 2$\,M$_\odot$ stars, finding 20 systems, 7 of which show tidally excited oscillations. The most massive found is HD\,5980 in the Large Magellanic Cloud, a young quadruple system with a total mass of 150\,M$_\odot$. They also found self-excited pulsations in 9 heartbeat stars with $\beta$\,Cep, SPB, $\delta$~Sct, and $\gamma$~Dor components. These will illuminate how tides interact with p\,modes, as previously observed in KIC\,4544587 \citep{2013MNRAS.434..925H}, but not yet probed for asteroseismic inference. 

\citet{2019ApJ...885...46G} studied the eccentric binary KIC\,4142768 which shows both self-excited g\,modes and p\,modes, as well as tidally excited modes. This combination can provide internal and surface rotation rates along with information on tidal dissipation, hence the timescale of circularization and synchronization in eccentric binary stars. Surprisingly, it seems that spin-orbit tidal interaction might possibly go either way in some stars. \citet{2020MNRAS.497.4363L} studied 45 non-eclipsing binaries found by the phase modulation (PM) method ({Section\,D.1 in the appendix}), for which 35 systems showed the g\,mode patterns needed to measure internal rotation. They found that, in general, tidal synchronism occurs for orbital periods less than 10\,d, with the exception of 3 stars for which an internal rotation rate was sub-synchronous. They conjectured that tidally excited modes may be able to transfer internal angular momentum to the orbit, in a novel process they dubbed inverse tides.  

\subsection{Tidally tilted pulsators} 
\label{ttp}

When I discovered the roAp stars and proposed the oblique pulsator model \citep[][Section\,\ref{roAp}]{1982MNRAS.200..807K}, the clue to oblique pulsation was in frequency multiplets split by exactly the rotation frequency. In slowly rotating stars, rotationally perturbed multiplets, such as dipole triplets and quadrupole quintuplets, have frequency splittings that differ from the rotation frequency by a factor of $(1 - C_{n,\ell})$, where $C_{n,\ell}$ is the Ledoux constant, which is not zero. The splitting of the multiplet components by exactly the rotation frequency showed that the mode axis was inclined to the rotation axis as a consequence of the strong, roughly dipolar magnetic field. This suggested that other perturbations from spherical symmetry should also compete with rotation for the pulsation axis selection, and that tidal deformation is an excellent candidate to do that. 

Forty years later, obliquely pulsating close binary stars have now been discovered where the pulsation axis coincides with the tidal axis. The first, HD\,74423 \citep{2020NatAs...4..684H}, is composed of two similar mass $\lambda$\,Boo stars (Section\,\ref{lambdaboo}) in a 1.6-d orbit where one star is a $\delta$~Sct star pulsating in a single mode which shows 11 frequencies in the FT split by exactly the orbital frequency -- the signature of oblique pulsation. A detailed analysis found a zonal ($m=0$) distorted dipole pulsation mode with much higher amplitude on the $L_1$ side facing the companion; this led to the star being called a single-sided pulsator. 

\begin{marginnote}[]
\entry{Tidally tilted pulsators} \ Obliquely pulsating stars in close binaries where the tidal axis is the pulsation axis. 
\entry{Single-sided pulsators} \ Tidally tilted pulsators where the tidal distortion leads to pulsation predominantly on either the $L_1$ or $L_3$ side.
\end{marginnote}

The second is CO~Cam \citep{2020MNRAS.494.5118K}, a marginal Am $\delta$~Sct star in a 1.27-d orbit with an undetected G main-sequence companion. It pulsates in four p\,modes with the pulsation axis aligned with the tidal axis, and three g\,modes that do not show this alignment. Pulsation modeling suggests that one of the p\,modes is the fundamental mode and that the others have a mixed p\,mode--g\,mode character. Mode identification independent of modeling is difficult because of the tidal distortion of the observable atmospheric layer. The best models require some core overshoot of 1\% to 2\% of the pressure scale height; models with no overshoot do not match the observed frequencies, hence there is an asteroseismic constraint on core mixing in this close tidal binary.  

The third is TIC\,63328020 \citep{2021MNRAS.503..254R}, a $\delta$~Sct star with a 1.1-M$_\odot$ companion in a 1.11-d orbit. There are two pulsation modes, where the primary one is a sectoral dipole ($\ell = 1, m = |1|$) mode. This star does not show the single-sided pulsation nature of the previous two, hence we refer to the class as tidally tilted pulsators with the term singled-sided pulsator reserved for those with much higher amplitudes on one side of the star -- either the $L_1$ or $L_3$ side. 

\citet{2020MNRAS.498.5730F} discussed the theory of tidal trapping, based on the above three tidally tilted pulsators. They examined how the tidal distortion affects the mode frequencies; how it causes alignment of the pulsation axis with the tidal axis; how it traps modes, largely confining them to the tidal poles or equator; and how it amplifies the mode amplitude at the tidal poles where the modes propagate closer to the stellar surface. Study of the orbital variations of HD\,74423, CO~Cam, and TIC\,63328020 showed that they partially fill their Roche lobes to varying degree, such that with only three examples it is not clear why their single-sided pulsations differ. 

The tidally tilted pulsators show asteroseismic promise. The oblique pulsation character helps with mode identification, a requisite for asteroseismology, even with the tidal distortion. The coupling of the surface p\,modes to interior g\,modes in a tidally distorted star requires further development of the models. The first three were discovered because their FTs are particularly simple, with only one or a few modes excited. Most $\delta$~Sct stars show many excited modes (Section\,\ref{deltasct}), and in the cases of tidally tilted pulsators with many modes, there is complexity to be disentangled in the FT. For relatively slowly rotating stars, the tool to do that is the \'echelle diagram, and it is probable that many more of these stars will emerge from the TESS data. Further developments to the theory are also needed, including incorporating the Coriolis and centrifugal forces, and possibly unanticipated surprises as this new behavior is explored. 

There is related behavior in other stars, e.g., the series of p\,modes separated by the orbital frequency, but which are not harmonics of that frequency, in the heartbeat star KIC\,4544587 \citep{2013MNRAS.434..925H}, and in the eclipsing binary U\,Gru \citep{2019ApJ...883L..26B}. For the latter, \citet{2019ApJ...883L..26B} discussed tidally excited and perturbed modes, and also single-sided mode trapping. It is clear that the interaction of pulsation modes and tidal distortion in close binaries offers new asteroseismic inference, with a range of behavior culminating in the tidally tilted pulsators. The problem of mode selection and excitation is not solved for these stars, but then, it is not solved for any $\delta$~Sct stars. At least in this case of close binaries, when the mode frequencies are harmonics of the orbital frequency, there is a clear coupling to the binarity that can be exploited. An observational goal is to find tidally tilted pulsators in eclipsing binaries where the eclipses add further measures of mass and radius, and the changes in the pulsation mode amplitude and phase during eclipse offers tomography of the mode geometry. 

\citet{2021MNRAS.501.1836Y} have theoretically studied the case of dynamical tides in close double degenerate white dwarfs with orbital periods less than an hour, in extreme cases less than 10\,min. These stars are, e.g., the double-degenerate interacting AM\,CVn stars, which are precursors of white dwarf mergers (Section\,\ref{whitedwarfs}), Supernovae Ia, and are a source of gravitational waves that LISA is expected to detect. \citeauthor{2021MNRAS.501.1836Y} predicted that coupling of g\,modes and p\,modes can produce photometric variations up to 1\%, which can be larger than the ellipsoidal variations. Furthermore, the pulsations generated by the dynamical tide have a phase shift of about $50^\circ$ to the orbital variation, which will allow them to be distinguished from the ellipsoidal variability. Even for faint white dwarfs, TESS observations offer the possibility to detect a 1\% effect, hence we can look forward to observations that will enable asteroseismic application of this new theory. 

\subsection*{Summary: Asteroseismology of close binary stars}
\begin{enumerate}
\item Eclipsing binaries with pulsating components provide multiple measures of stellar fundamental parameters. Together they constrain mass, radius, age, and luminosity to the highest precision for application in exoplanet and galactic archeology studies. 
\item The phase modulation (PM) technique has opened up unexplored parameter space in the orbital periods of binary stars. It allows orbital solutions for binary stars without the need of spectroscopy to obtain radial velocities, saving vast amounts of telescope time and astronomers' time.
\item Doppler beaming also allows the measurement of radial velocities without the need of spectra. 
\item Heartbeat stars and tidally tilted pulsators provide observations to test theories of stellar tides and their interaction with pulsation, and with rotational and orbital angular momentum. They test mode trapping, tidal distortion of pulsation frequency and amplitude, and pulsation mode axis selection.
\item The heartbeat stars support the thesis of tidal locking where feedback among orbital, rotational, and stellar evolution changes keep the pulsation frequency resonant with the orbital frequency.
\end{enumerate}

\section{Epilogue -- past and future}

I am fortunate that my career has spanned the time from the beginning of helioseismology in the 1960s through to this blossoming of asteroseismology today. This review has highlighted the results for stars across the HR~Diagram (except for the solar-like oscillators and red giants, as a consequence of other recent reviews of those iconic asteroseismic targets). All of this work uses asteroseismology to  see inside the stars. Fundamental and other stellar parameters are extracted from asteroseismic inference: mass, radius, age, metallicity, atomic diffusion, luminosity, distance, magnetic fields, interior rotation, angular momentum transfer, convective overshoot, core burning stage. In the spirit of Howard Florey, asteroseismologists are delighted to be exploiting stellar pulsation in more depth than previously possible, and to be discovering new kinds of pulsation behavior, all improving our understanding of stellar structure and evolution, a bedrock of astrophysics, since stars are the luminous tracers of the universe. 

Space photometry and ground-based large scale surveys have created an explosion of asteroseismic data. Big data techniques and machine learning are already active in exploring and categorizing the huge data sets, and this will expand in the future. Human labor cannot sift the large numbers of light curves and FTs as computers can. This is particularly true of the ground-based surveys such as ZTF and VRO, the latter of which will produce 20\,Tb of data per night! These ground-based surveys will enable and expand asteroseismology of many classes of stars discussed in this review, such as white dwarfs, Blue Large Amplitude Pulsators, EL\,CVn binaries, Cepheids, RR\,Lyr stars, and Stochastic Low Frequency O and B stars. 

The photometric space missions have been designed for exoplanet studies, and they have been successful beyond expectation, with exoplanets now known in their thousands, soon to be tens of thousands. For these missions we asteroseismologists initially rode on the coattails of the exoplanet community, but now that the usefulness of asteroseismology to exoplanets is known, asteroseismology is built in to the science case for future missions. Nevertheless, hot stars are not the main hosts of exoplanets, and they are not expected to have inhabited planets, even if habitable planets orbiting them should be found, for the obvious reason of their short lifetimes. Hot main-sequence stars are also rarer as a consequence of the initial mass function and those short lifetimes. But, as this review has shown, the massive upper main-sequence stars, the white dwarfs and subdwarfs, the most luminous blue giants and supergiants, and the classical pulsators, are all of interest asteroseismically. 

Most of the variability in hot pulsators comes from variations in $T_{\rm eff}$. As a simple consequence of their spectral energy distributions, the hot stars have much larger pulsation amplitudes in the UV and Blue. This is easy to see by picturing the changes in a blackbody curve with changes in temperature for stars for which the flux peaks in the UV or blue. The TESS bandpass is roughly $0.6 - 1.0$\,$\upmu {m}$, that of {\it Kepler} was $0.4 - 0.9$\,$\upmu {m}$, filters with broadly white bandpasses, but towards the redder end of the spectrum. That is the right choice for the study of exoplanet transits and of the stochastic pulsators such as the solar-like stars and red giants. But to obtain much higher signal-to-noise for the hotter stars, a blue filter is needed. A dedicated space mission, which could be called the Blue Asteroseismic Survey Satellite (BASS), is needed to build on the revolutionary asteroseismic results obtained from the exoplanet surveys. 

Asteroseismology informs our understanding in hotter stars: 

\begin{itemize} 
\item of stellar structure and evolution, including interior rotation, angular momentum transfer, and mixing;
\item of the most massive stars, precursors to core-collapse supernovae, black holes and neutron stars, gamma ray bursters, observable gravitational wave events, and sources of much of the chemical evolution of the galaxy.
\end{itemize}

in cooler stars:  
\begin{itemize}
\item of galactic archeology; 
\item of the fundamental parameters -- mass, radius, and age --  needed to characterize exoplanet host stars in our search for habitable planets;
\item of red giant evolutionary core-burning state;
\item of the post-main-sequence giant to red giant stages, including interior rotation and magnetic fields. 
\end{itemize}

in compact stars:
\begin{itemize}
\item of mass, radius, temperature, luminosity, age, rotation, the extent of the non-degenerate surface layers in white dwarfs;
\item of physics under extreme conditions in white dwarfs and subdwarfs, the end stages of the evolution of 97\% of all stars; 
\item of double and single degenerate binary stars, precursors of distance scale Supernovae\,Ia, and further sources of chemical evolution. 
\end{itemize}

in close binary stars:
\begin{itemize}
\item of tides, tidal distortion of pulsation modes, and of mode trapping;
\item of orbital and rotational angular momentum evolution and exchange; 
\item of orbital circularization. 
\end{itemize}

\noindent We have indeed found Eddington's ``appliance'' to ``pierce the outer layers of a star and test the conditions within''.  He would have been  amazed and gratified, as are we.

\section*{DISCLOSURE STATEMENT}
The author is not aware of any affiliations, memberships, funding, or
financial holdings that might be perceived as affecting the objectivity of this
review.

\section*{ACKNOWLEDGMENTS}

I thank Professors Conny Aerts and Rob Kennicutt for their many scientific comments and their guidance in the preparation of this review. I also thank Professor Simon Jeffery for providing and discussing the pulsation HR~Diagram shown in Figure 1, modified for this review. 

\addtocontents{toc}{\protect\setcounter{tocdepth}{1}}


\section{Appendix: Introduction to the supplementary materials}

In writing this review there were many personal stories of my own and historical stories of other researchers that I wished to tell. Some of those are in the main text, and others are here in this supplement. These stories give a sense of the human side to the research, especially for those entering the field and for those from other fields who are reading the review out of general interest. In addition, for many sections of the review I wanted to provide more depth for the specialist. Those additional details are in the sections of this supplement. The review stands on its own. For those who enjoy the stories and want the additional details, I recommend when reading the main text that you take an excursion to this supplement directly from the links provided.  

\appendix

\section{OBAF stars}

\subsection{The naming of the SPB stars and discovery of the pulsation driving mechanism in B stars}
\label{app:SPB}

\subsubsection{The naming of the SPB stars} At a stellar pulsation meeting in 1990 in Bologna, Italy, Christoffel Waelkens gave a talk about a new class of pulsating late-B stars that he proposed to name the Slowly Oscillating B stars, in analogy with the rapidly oscillating Ap stars that I had previously named. In the question time at the end of his talk, I commented that his class would come to be known by its acronym, and that he really did not want to be studying SOBs. He and some of the audience expressed some confusion about this English colloquialism, but after it was discreetly explained, he changed the name to Slowly Pulsating B (SPB) stars \citep{1990ASPC...11..258W,1991A&A...246..453W}.

\subsubsection{The pulsation driving mechanism in B stars} When I was a student at the University of Texas in the early 1970s, I presented my first professional talk at the December 1973 AAS meeting in Tucson, Arizona. It was a 10-min talk on the evolved metallic-lined stars, the $\delta$\,Del stars (now called $\rho$\,Pup stars) \citep{1976ApJS...32..651K}. My talk was in a session where all other talks were about numerous ideas to explain pulsation driving in $\beta$\,Cep stars -- then a mystery. Not one of those ideas was conclusive, and, as it turned out later, not one was correct. At the end of my talk there were no questions from the audience, so the chair of the session, Martin Schwarzschild, asked forcefully and loudly, in his way, ``WHAT are these stars REASONS to PULSATE?'' I thought I knew the answer to his question, but since one of my undergraduate textbooks had been his {\it Structure and Evolution of the Stars} \citep{1958ses..book.....S}, I thought to myself, ``if he doesn't know, then what I am thinking cannot be right''. He terrified me, so I diverted the question to the theoreticians in the audience who had given the talks on $\beta$\,Cep driving mechanisms. One of them gently answered, ``Martin, they are in the instability strip'', to which he softly replied, ``Oh. I missed that.'' I understood, on meeting him in later years, that he was just being kind in providing a poor graduate student giving his first professional talk with a question. 

The problem of driving in $\beta$\,Cep and SPB stars was that the $\kappa$-mechanism needs abundant ions to provide sufficient driving to overcome damping in other layers of a star, and the prime drivers for many pulsating stars are H and He, the most abundant elements (as Cecilia  Payne had shown). H and He do not work for the B stars as their ionization zones, where the driving occurs, are too close to the surface of the star, so that there is insufficient mass in overlaying layers to cause recompression. \citet{1992ApJ...393..272C} and \citet{1992A&A...256L...5M} were among the first to show that a bump in opacity from heavy elements -- the Z-bump -- primarily from Fe and Ni, could drive pulsations among the B stars. After decades, the driving mechanism for these hot main-sequence stars had been found. See \citet{1996ARA&A..34..551G}, \citet{1999LNP...523..320P} and \citet{2007MNRAS.375L..21M} for further discussion of the driving mechanism and the instability strips of B stars. 

\subsection{The discovery of the $\gamma$~Dor stars}
\label{app:gammador}

Michel Breger was a pioneer in the study of $\delta$~Sct stars (e.g., \citealt{1969ApJS...19...79B}), and under his supervision I also began studying them in the early 1970s. Using ground-based differential photometry, we saw low frequency excess in the amplitude spectra of the stars we studied. Because we were using at least two comparison stars for each $\delta$~Sct star studied, we did not understand the origin of this low-frequency excess, and dismissed it as probably an artifact caused by differential extinction, given that our comparison stars had different temperatures to the $\delta$~Sct stars. We know now that he and I were probably seeing evidence of g\,modes in $\delta$~Sct stars, but we did not recognize this at the time.

The history of the discovery of g\,mode pulsation in A and F stars is interesting. Variability in $\gamma$~Doradus itself was first noted by \citet{1963MNSSA..22...65C}\footnote{Alan Cousins published in MNRAS from 1923 to 2001, the year of his death. This 74-year publishing span may be a record.} who had used the star as a comparison star for $\beta$\,Dor (a Cepheid), but noticed an excess scatter in the $\gamma$~Dor measurements. \citet{1989IBVS.3412....1C} then noted that $\gamma$~Dor has a period close to 0.74\,d. \citet{1992Obs...112...53C} reported the analysis of data gathered over years at the South African Astronomical Observatory (SAAO) by Fred Marang, Francois van Wyk, Bob Stobie, and himself. It was not the custom at that time to include these others as co-authors, although they were acknowledged. Alan Cousins spoke independently to Luis Balona and to me about the frequency analysis of his data. I failed to disentangle the amplitude spectra, but Luis Balona did so successfully, and was acknowledged in Cousins' paper. \citet{1992Obs...112...53C} announced that $\gamma$~Dor has two pulsation frequencies, $1.32098$\,d$^{-1}$ and $1.36354$\,d$^{-1}$. He noted that these are too low in frequency for a $\delta$~Sct star, and he speculated about possible spots or tidal distortion as an explanation for the two frequencies. It was not understood at that time that these two frequencies are from g\,modes. 

TESS has now observed $\gamma$~Doradus itself in 6 Sectors,  and I have analyzed those data to compare with the early work. Stunningly, there are two principal pulsation modes in the TESS data with frequencies of $1.32098$\,d$^{-1}$ and $1.36378$\,d$^{-1}$, the same as Cousins' values within the uncertainties. That is remarkable, given that he and his collaborators were working with sparse ground-based data spanning years, giving amplitude spectra with complex, and confusing, spectral windows. In a further study of $\gamma$~Dor, \citet{1994MNRAS.267..103B} correctly suggested that $\gamma$~Dor may be an example of a new class of pulsating star, then \citet{1994MNRAS.270..905B} suggested that $\gamma$~Dor is the best example of that new class. That was the discovery paper of the class of $\gamma$~Dor stars, a class that was later defined by \citet{1999PASP..111..840K}.

\subsection{Frequency variability in two $\delta$~Sct stars}
\label{app:freqvar}

It is not surprising to find frequency variability in $\delta$~Sct stars, given their complex behavior with many modes, both radial and nonradial, simultaneously excited. This is illustrated here with two, of many, cases: that of 4\,CVn for which \citet{2017A&A...599A.116B} analyzed over 800 nights of ground-based photometry obtained over more than 40\,yr; and that of KIC\,5950759 \citep{2021MNRAS.504.4039B} for which 4\,yr of {\it Kepler} data were analyzed. 

\citet{2017A&A...599A.116B} noted the complexity of the FT of 4\,CVn and concentrated their attention on higher amplitude modes. They found both frequency and amplitude variations that are correlated from mode-to-mode over the decades, although the signs and amplitudes of the variations differ. The dominant modes studied were nonradial dipole and quadrupole modes for which, interestingly, they found frequency changes of opposite sign for modes of order $+m$ and $-m$. Since the frequency separation of those modes is a measure of rotation in the pulsation cavity, their result suggests internal changes in the rotation of 4\,CVn over the 40\,yr of observations. 

\citet{2021MNRAS.504.4039B} combined photometry for KIC\,5950759 from both ground-based WASP observations and a 4-yr {\it Kepler} data set, along with spectroscopy to determine fundamental parameters. KIC\,5950759 is unusual: it is a high amplitude $\delta$~Sct (HADS) star (meaning $\Delta V > 0.3$\,mag peak-to-peak), and it has low metallicity, hence may also be an SX\,Phe star (Population\,II $\delta$~Sct star). Its FT shows an abundance of harmonics and combination frequencies for the two highest amplitude frequencies, which were identified as the radial fundamental and first-overtone modes. Evolutionary models put the star in the contraction phase at the end of the terminal-age main-sequence (TAMS), but the measured period changes for both modes, $\dot  P/P = 10^{-6}$\,yr$^{-1}$, are two orders of magnitude higher than expected from evolution. 

Asteroseismic frequency changes have great promise for measuring stellar evolution in real time. But we must identify and separate all the cases of frequency changes arising from other causes to realize that promise. 

\subsection{New techniques to determine internal rotation}
\label{app:internalrot}

\citet{2020A&A...635A.106T} developed a new, model-independent diagnostic of internal rotation in $\gamma$~Dor stars using prograde sectoral g\,modes. Instead of modeling the slopes in the $\Delta P - P$ diagram, they plotted the pulsation frequencies, $\nu$, versus the square root of the frequency differences, $\sqrt{\Delta \nu}$, then used that to make mode identification, determine the average internal rotation rate, and measure $\Pi_0$ ($P_0$ in their notation). In a test of a few cases, they showed that the rotation rates are consistent with those found for the same stars previously in multiple other studies. This new method will be useful for large ensembles of $\gamma$~Dor and SPB stars expected from TESS data where series of prograde sectoral g\,modes frequencies are found. Following \citet{2016A&A...593A.120V}, \citeauthor{2020A&A...644A.138T} extended the method to model r~modes, too, with improved estimates of the average internal rotation and of $\Pi_0$. 

In an interesting development, \citet{2020A&A...640A..49O} and \citet{2021MNRAS.502.5856S} independently examined the interaction of core inertial modes with the surrounding radiative envelope. Inertial modes occur where the Coriolis force is primary restoring force; they have frequencies between zero and twice the rotation frequency of the cavity in which they are trapped, in this case the stellar core. These modes in the rotating convective core can resonantly couple with g\,modes (hence gravito-inertial modes) propagating above that core. This gives an observational measure of the rotation of the convective core, something the g\,modes by themselves cannot provide, as they are evanescent in convection zones. 

\citeauthor{2020A&A...640A..49O} provided a theoretical exposition for this new connection, and \citeauthor{2021MNRAS.502.5856S} applied the method to 16 $\gamma$~Dor stars.  They found a distinctive dip in the $\Delta P - P$ diagram, which occurs where there is a resonance at a spin parameter $2f/\nu \sim 8 - 11$, where $f$ is the rotation frequency of the convective core and $\nu$ is a g\,mode pulsation frequency in the co-rotating frame. Allowing the core to rotate differentially to the surrounding envelope in their models, they found for the 16 stars nearly uniform rotation, with less evolved stars having slightly higher core rotation rates. \citeauthor{2020A&A...640A..49O} found in their models multiple resonances possible, although only one dip in the the $\Delta P - P$ diagram has yet been observed for each star. 
 
In light of these results, note that KIC\,11145123 \citep[][Section\,2.4]{2019ApJ...871..135H}, with a core rotation rate 6 times the envelope rate, differs from the 16 $\gamma$~Dor stars studied by \citet{2021MNRAS.502.5856S}. However, KIC\,11145123 rotates very slowly; even at 6 times the envelope rotation rate the core rotation period is still only $\sim$17\,d, whereas all 16 stars studied by \citeauthor{2021MNRAS.502.5856S} have core rotation periods in the $0.4 - 0.8$\,d range. 

\subsection{Two examples of atomic diffusion in asteroseismic models.}

\citet{2017A&A...601A.127D} first noted the importance of including atomic diffusion in asteroseismic modeling of exoplanet host stars, which they found increased the age determined in the exoplanet host star 94\,Cet by few percent. They noted that the effect on age could be larger for hotter stars and for lower metallicity. It is clear that the inclusion of atomic diffusion for asteroseismic modeling is important for stars with masses greater than 1.4\,M$_\odot$.

\citet{2020ApJ...895...51M} added atomic diffusion to models for asteroseismic inference. They found a better match to the observations in the case of the $\delta$~Sct star KIC\,11145123 (Section\,2.4) with atomic diffusion than without it, although in another case the atomic diffusion did not provide better models. For KIC\,11145123 they also examined whether atomic diffusion might explain its abnormal abundances. Spectroscopic analysis by \citet{2017MNRAS.470.4908T} found that the star is metal weak, $[{\rm Fe/H}] = -0.71$. From its high radial velocity and position above the galactic plane, they suggested that it is a blue straggler that has undergone mass accretion from a companion. \citet{2020ApJ...895...51M} revisited this problem, and using distance and proper motion from Gaia found the star to be 260\,pc away from the galactic plane with a peculiar velocity of 72\,km~s$^{-1}$, suggesting it might be a runaway star. They attempted to model the surface abundances with atomic diffusion and had some success, but could not reproduce the metal deficiency starting with a solar abundance model. KIC\,11145123 still has mysteries to solve.

\subsection{Evidence of weak spots in upper main-sequence stars}
\label{app:spots}

\citet{2011A&A...536A..82D} presented evidence of photometric variation due to spots in a B star, HD\,174648, observed by CoRoT. From a sample of about 90\,000 stars in the {\it Kepler}, K2 and TESS data, \citet{2021FrASS...8...32B} suggested that about $30 - 40$\% of upper main-sequence stars in the {\it Kepler} data show evidence of rotational variation. \citet{2020MNRAS.491.3586L} examined a sample of 611 $\gamma$~Dor stars, and,  using a more stringent criterion than \citeauthor{2021FrASS...8...32B}, found 9.5\% of those stars to have rotational variations. The origin of spots in upper main-sequence stars is not yet known. \citet{2019ApJ...883..106C} examined the possibility that weak magnetic fields of order of a few Gauss could be generated by dynamos in the surface convection zones of A stars; perhaps those could generate spots. Whatever the resolution of this problem of origin, the evidence is good that there are weak spots (compared to the Sun) that cause rotational light variations in upper main-sequence stars. Since the rotation periods for many of these stars are in the same range as the g\,mode and r~mode pulsation periods, care is necessary to identify correctly  the surface rotation signal. The importance of doing so is that the spots and r~modes measure the surface rotation period, which can then be compared to the deep near-core period obtainable from asteroseismology using the g\,modes. \citet{2020MNRAS.491.3586L} did just that, using period spacings of g\,modes to examine near core rotation rates, and spot variations, where identified, to measure surface rotation rates. 

\subsection{Be star history}
\label{Be-appendix}

Angelo Secchi ($1818 - 1878$) was an Italian Jesuit priest and astronomer. Following the Roman revolution of 1848, he spent two years at Stonyhurst College in Lancashire -- not far from my own UK institution -- before becoming the director of the Roman College Observatory on his return in 1850, a position he held until his death in 1878. In 1870 when Italy was unified and the final papal state, which included Rome, was incorporated in the Kingdom of Italy, Secchi refused to shift his allegiance (of course!) from Pope Pius IX to King Victor Emmanuel II. He was illustrious by then and remained director of the observatory by the expedient of threatening to leave Italy and take up one of offers he had elsewhere in Europe. The Royal government chose not to interfere. The observatory was nationalized on Secchi's death in 1878, but was resurrected as the Vatican Observatory in 1891. Secchi is most famous for his work in spectroscopy and his first introduction of his five spectral classes, later superseded by the Harvard classes we use now. One of his first discoveries after installing a new spectroscope was finding that $\gamma$\,Cas (classified B0IVe now) had a ``very beautiful luminous line''  \citep{1866AN.....68...63S} which he identified as the Fraunhofer F-line (H$\beta$); he further noted that in addition to that ``shiny band'' there are others. He concluded that ``This star therefore has a spectrum opposite that of the ordinary type of white stars.'' This was the discovery of emission lines in stars, and, in particular, was the first B emission line star of a class now known as Be stars. 

\subsection{More about roAp stars}
\label{app:roap}

\citet{2007A&A...473..907R} derived the variation of pulsation amplitude and phase in 10 roAp stars from line bisector studies, showing how the pulsation amplitude rises with atmospheric height, accompanied by pulsation phase changes.  \citet{2009MNRAS.396..325F} studied the radial velocity pulsation and rotation variations in the singly-periodic roAp star HD\,99563, with contemporaneous data from the 8-m VLT and Subaru telescopes obtained over the full 2.91-d rotation cycle. From a study of spectral lines of Ca, Eu, Nd, and the core of H$\alpha$, they found pulsation amplitude increasing with atmospheric height, also accompanied by phase changes. They found that the single pulsation mode is a distorted dipole mode, and that the distortion appears to be different for spectral lines of different ions, meaning that the geometry of the pulsation mode varies with atmospheric depth, consistent with the findings of  \citet{2006A&A...446.1051K} for HR\,3831 and \citet{2020ASSP...57..313K} for  HD\,6532.

\citet{2021A&A...650A.125D} examined the fundamental properties of 14 roAp stars using both interferometry and asteroseismology with a grid of models.  Interestingly, they found that roAp stars are systematically less massive and older than non-pulsating Ap stars, a result that may inform the still problematic question of pulsation driving in these stars. They also constrained their models better with the additional knowledge of the large separation, $\Delta \nu$, for two stars, HR\,1217 and $\alpha$\,Cir. While the magnetic field in roAp stars significantly perturbs the mode frequencies, it does so almost equally for all modes, so that the large separation is still an important asteroseismic constraint. \citeauthor{2021A&A...650A.125D} found normal, solar metallicities for those two stars, which when contrasted with the stunning overabundances of the rare earth elements in the atmosphere is an important observational constraint on atomic diffusion. 

Detailed asteroseismology of the deep interior of roAp stars will be difficult. The high overtone magneto-acoustic modes are most sensitive to the outer envelope, and the gravito-inertial g\,modes that so beautifully illuminate our view of the cores of $\gamma$~Dor and SPB stars (section\,2.5) are not observed. \citet{2020MNRAS.498.4272M} found in magnetic models that some low overtone g\,modes can be excited in Ap stars, but high-overtone g\,modes are heavily damped in stars with field strengths stronger than $1-4$\,kG, as is typical in Ap stars. They also showed from models that field strengths greater than about 1\,kG strongly damp low-overtone $\delta$~Sct p\,modes, which, if they were observed,  could inform the conditions somewhat deeper than the high-overtone modes typically observed. 

\subsubsection{Frequency variability in roAp stars} 
\citet{2021FrASS...8...31H} reviewed {\it Kepler} observations of roAp stars. Only 14 roAp stars -- both known and newly discovered -- were observed by {\it Kepler} and all show some frequency variability (as do many $\delta$~Sct stars, Section\,2.2). That frequency variability is typically non-cyclic, as is the case for 16\,yr of ground-based observations of the roAp star HR\,3831 \citep{1997MNRAS.287...69K}. As with frequency variability in many classes of pulsating stars, this may be evolutionary weather (Section\,2.2). Ap stars in general are seldom found in close binaries \citep{2012A&A...545A..38S}, so the frequency variability does not have a contribution from orbital motion (Section\,D.1). 

\subsubsection{TESS and roAp stars} 

TESS observed 960 stars classified as Ap in 2-min cadence in all 13 sectors of cycle 1 to determine the fraction of Ap stars that are roAp stars, and to find new roAp stars for further analysis. In addition, a team searched the light curves of over 50\,000 stars hotter than $T_{\rm eff} = 6000$\,K for the signature of roAp pulsation in the FTs \citep{2021MNRAS.506.1073H}. Together with earlier examinations of Ap stars in subsets of TESS cycle 1 sectors by \citet{2019MNRAS.487.2117B} and \citet{2019MNRAS.487.3523C},  \citet{2021MNRAS.506.1073H} found 12 new roAp stars and further analyzed 44 previously known ones. There are, as of their writing, 88 known roAp stars. It had previously been suspected that roAp stars only appeared rare because their pulsation amplitudes were too low to be detected in ground-based observations. The TESS data show that they are truly rare, finding only 3\% of the Ap stars observed to be roAp stars. The reason for this low incidence is not known.

The TESS data yielded the longest pulsation period in an roAp star, 25.8\,min, and the shortest, 4.7\,min, extending the period range, hence radial overtones for these stars. The relation between rotation period and roAp pulsations is not yet known, hence the increased sample provides more data to examine this.  \citet{2021MNRAS.506.1073H} found that 68\% of the 56 roAp stars with TESS data show rotational modulation from spots. This is useful for precise determination of the rotation period, but it is only for stars with $P_{\rm rot} \simkl 22$\,d, as a consequence of the 27-d TESS sectors. 

\citet{2020A&A...639A..31M} hypothesized, then showed that Ap stars with TESS data that do not show rotational modulation due to spots are almost all super-slowly-rotating Ap (ssrAp) stars, which they defined to be magnetic Ap stars with rotation periods longer than 50\,d. Of the 54 ssrAp stars they found in cycle 1 data and 68 in cycle 2 data (Mathys, Kurtz \& Holdsworth, in press), at most a few can be pole-on, hence they significantly increased the number of ssrAp stars known. These stars have magnetic field strengths in the range $0 - 10$\,kG typical of Ap stars, and they have rotation periods known, in some cases, to be greater than a century. Hence they address the question of magnetic braking. Interestingly, 22\% of 54 ssrAp stars in cycle 1, and 19\% of the 69 ssrAp stars in cycle 2, are roAp stars (or candidates), an apparently significantly larger fraction than the 3\% found for the whole 960 Ap stars sample. This thus begins to address the relation between rotation and pulsation in roAp stars. 

\begin{marginnote}[]
\entry{ssrAp stars} \  Super slowly rotating Ap stars with rotation periods greater than 50\,d up to centuries. 
\end{marginnote}

A prime reason for finding more roAp stars is to find stars with rich FTs with alternating even and odd degree modes that readily give mode identification for asteroseismic modeling. The finding of new roAp stars also leads to objects with higher amplitudes suitable for future high time resolution spectroscopic study  with large telescopes.  \citet{2021MNRAS.506.1073H} provide first analyses of the roAp stars with TESS data, characterizing their pulsation and rotation. As in the study of HD\,86181 \citep{2021MNRAS.tmp.1926S}, many stars have rotationally split multiplets useful for constraints on mode geometry, and some stars have multiple modes suitable for future asteroseismic modeling to determine fundamental parameters such as mass, age, luminosity, and overall metallicity in the pulsation cavity. Determining the latter may inform the question of the depth of the extreme peculiarities seen in the atmosphere, and that is relevant to theoretical calculations of atomic diffusion timescales. 

\section{Pulsating white dwarf stars}
\label{app:whitedwarfs}

\subsection{The Chandrasekhar -- Eddington debate}
\label{app:whitedwarfs-chandra}

It is less than 100 years ago that in 1930 Subrahmanyan Chandrasekhar arrived by ship from Chennai (then Madras) to begin his postgraduate studies at Cambridge University. On the sea voyage he spent his time calculating the maximum mass a white dwarf star can have considering the quantum mechanical Pauli exclusion principle and including, for the first time, Special Relativity. He was a 20-year-old prodigy at the time. During the early 1930s Chandra derived the complete solution to the problem of a relativistically degenerate star. He famously presented this work publicly at the Royal Astronomical Society January 1935 meeting at Burlington House. He was surprised to have his work dismissed in a following talk by Professor Arthur Eddington (whose quote from {\it The Internal Constitution of the Stars} opened this review). Eddington did not believe Chandrasekhar's results on white dwarf stars, and such was Eddington's stature that Chandrasekhar's result was not widely accepted by astronomers for decades. It was, however, accepted at the time by physicists, notably Niels Bohr and Wolfgang Pauli. While Eddington and Chandrasekhar had a generally friendly personal relationship, the acrimonious debate on the nature of white dwarfs appeared to have overtones: of age and authority (Professor -- student), and race and nationality (it was a time of Empire); see  \citet{1982PhT....35j..33W} for a thorough discussion of this confrontation. Chandrasekhar was worn down by the debate, when he knew his results were correct, so he published a textbook  \citep{1939isss.book.....C} on stellar structure (which I used as an undergraduate and still have on my bookshelf) and moved on to other fields. His recognition for the derivation of the Chandrasekhar limit of $1.44$\,M$_\odot$ for the maximum mass of a white dwarf was awarded the Nobel Prize in Physics in 1983. 

Eddington believed that all stars must end their lives as white dwarfs, hence his antipathy to Chandrasekhar's mass limit. In the 1930s the two other end states for stars, neutron stars and black holes, were yet to be understood, and it was only in the mid-1960s  to early 1970s  that neutron stars and black holes were observationally discovered. Initially, it was thought that only stars with masses less than $1.44$\,M$_\odot$ could become white dwarfs, but we now know that mass loss -- primarily in winds and shell loss episodes during late thermal pulses of Asymptotic Giant Branch stars, but also by other mechanisms -- allows about 97\% of all stars to end up as white dwarfs as their final stages. So Eddington's visceral desire that all stars end as white dwarfs was almost correct, albeit for wrong reasons and incorrect arguments. Chandrasekhar was right.  The 3\% of stars that Eddington was not correct about produce the spectacular physics: of neutron stars;  of stellar-mass black holes -- the seeds of supermassive black holes, hence galactic jets and quasars; and of merging neutron stars and black holes -- the sources of detectable gravitational waves.  

\subsection{ZZ\,Cet stars (DAV)}
\label{zzcet}

The vast majority of white dwarfs show H in their atmospheres, hence are spectroscopically classified DA. Their temperatures range from 170\,000\,K down to around 4500\,K for the coolest white dwarfs known, where that lower limit is set by the age of the galaxy; the first white dwarfs that formed when the galaxy was young have not yet had time to cool any further.  Asteroseismically, the most abundant pulsating white dwarfs are the DAV ZZ\,Cet stars. If we could observe white dwarfs to the distances we can for brighter stars, white dwarfs would be the most abundant pulsating stars observed in the galaxy. 

It is now more than 50 years since the discovery of the first pulsating white dwarf star, HL\,Tau\,76 \citep{1968ApJ...153..151L}. In that time there have been excellent asteroseismic successes in the studies of these stars. Particularly, the spacing of the g\,mode periods in white dwarfs yield a precise mass measure. A mass-radius relation then leads to a radius determination, somewhat dependent on the metallicity. Forward modeling leads to asteroseismic luminosity and temperature estimates. Given radius, temperature, and luminosity, an asteroseismic distance can be calculated and compared with Gaia distances, with good agreement. 

\begin{marginnote}[]
\entry{ZZ\,Cet stars} \ DAV white dwarfs; g\,mode pulsators.  $T_{\rm eff}$ of $10\,400 - 12\,400$\,K;  periods of $100- 1400$\,s. Typical mass is around $0.6$\,M$_\odot$.
\end{marginnote}

Deviations from regularity in the period spacing in white dwarfs give measures of the extent of the outer, non-degenerate layers where the modes are trapped. The precise period spacings also give constraints on atomic diffusion (Section\,2.7), an important mechanism of stratification in white dwarfs. The primary asteroseismic data in white dwarfs are, therefore, g\,mode period spacings with mode identifications. A robust method to identify those modes comes from recognition of rotationally-split multiplets -- typically triplets for dipole modes and quintuplets for quadrupole modes. A collateral, important benefit is the ability to determine the rotation period, hence angular momentum for the pulsation cavity, the outer envelope. 

White dwarf studies continue from the ground, outpacing what has yet been accomplished with {\it Kepler} and TESS data. An outstanding new initiative is the exploitation of data from the Zwicky Transient Facility (ZTF). The ZTF \citep{2019PASP..131a8002B} is a northern analogue to the Vera Rubin Observatory (VRO), albeit in a much smaller telescope (1-m aperture, as opposed to 8-m). Among the numerous, broad science objectives of the ZTF are stellar variability studies \citep[section 8]{2019PASP..131g8001G}. With a shortest cadence of 38.3\,s, the ZTF will discover abundant pulsating white dwarfs, the majority of which will be DAV stars. It will, of course, find vast numbers of other types of pulsating stars -- possibly mostly $\delta$~Sct stars -- from which automated searches will seek, find, and classify objects for followup, much of which will be ground-based, but some of which will be with TESS, PLATO, and their successors in space.

\begin{marginnote}[]
\entry{ZTF} \  Zwicky Transient Facility
\entry{VRO} \  Vera Rubin Observatory; previously known as the Large Synoptic Survey Telescope (LSST)
\end{marginnote}

An stunning example of this potential is the discovery of a 6.94-min rotation period in ZTF\,J190132.9+145808.7, a white dwarf with a mass of nearly the Chandrasekhar mass so that it almost as small as the Earth's Moon ($R \sim 2100$\,km), with a magnetic field in the $600 - 900$\,MG range \citep{2021arXiv210708458C}. This star is probably the result of a merger of two white dwarfs with orbital angular momentum being transferred to rotation in the remnant, hence the very short rotation period. This is not the shortest known white dwarf rotation period, however. \citet{2020ApJ...894...19R} discovered in {\it Kepler} data the fastest rotating, isolated white dwarf star, EPIC\,228939929. This star has a rotation period of 5.28\,min seen in rotational light variations caused by a spot associated with a 5\,MG magnetic field.

The ZTF discovered ZTF\,J190132.9+145808.7, and ground-based followup, both photometric and spectroscopic, led to the discoveries and conclusions discussed above. Large Scale Surveys have been spectacularly successful in pulsating star research, OGLE and ASAS being prime examples. But those did not have the short cadence capability of the ZTF, hence that facility leads in the detection of pulsating, rotating and eclipsing compact stars, where the periods are as short as minutes. TESS will provide long duration, precise light curves for some of these objects with a 20-s cadence, in the best cases, and 2-min cadence in many others.  

\citet{2017ApJS..232...23H} studied 27 ZZ\,Cet stars observed by the {\it Kepler} K2 mission, with spectra obtained for all 27. They discovered that pulsation periods shorter than about 800\,s are stable over the 90-d time span of the K2 data, but periods longer than 800\,s have a Lorentzian profile in the FT reminiscent of the stochastic pulsators, such as solar-like stars and red giants, implying modes lifetimes of only a few days to weeks, similar to the theoretical damping times for longer period g\,modes in white dwarfs. This is an interesting new discovery that is yet to be understood or exploited. \citeauthor{2017ApJS..232...23H} measured the internal rotation of 20 of the 27 stars from the frequency splitting of dipole triplets, doubling the number of DAV stars for which asteroseismic rotation rates are known.  The rotation periods are as short as hours, but typically just over a day. This is far slower than would be the case if the red giant core conserved angular momentum as it contracted with evolution, supporting strong angular momentum transfer between core and envelope in red giant stars. Asteroseismology of mixed modes in red giants show their cores to be rotating about 10 times faster than their envelopes (\citealt{2012Natur.481...55B}; \citealt{2012A&A...548A..10M}), giving periods in agreement with the white dwarf rotation periods \citep{2019ARA&A..57...35A}. 

\citeauthor{2017ApJS..232...23H} also showed that the DAV stars observed by {\it Kepler} have the same correlation of pulsation period with temperature previously known from ground-based studies: the weighted mean pulsation period (WMP) decreases from the blue to red edges of the instability strip. The WMP (which could have been called $P_{\rm max}$) is similar to $\nu_{\rm max}$, the frequency of maximum pulsation amplitude in stochastic pulsators. In red giants $\nu_{\rm max}$ decreases with increasing luminosity because of  decreasing sound speed and increasing radius (hence the Leavitt Law -- the PL relation); in DAV white dwarfs $P_{\rm max}$ increases with decreasing temperature because the surface convection zone deepens as the stars cool. 

Pulsation in DAV stars is driven by convection in the surface H ionization zone (\citealt{1991MNRAS.251..673B}; \citealt{1999ApJ...519..783W}). As white dwarfs cool and cross the ZZ\,Cet instability strip at around $13\,000$\,K (depending on mass), excitation in the relatively thin convection zone drives pulsations in low radial overtone modes (of which the dipole and quadrupole modes are observable) with periods in the $100 - 300$\,s range; the mode frequencies for these are stable. Since white dwarfs evolve simply by cooling, as the star cools across the instability strip, the surface convection zone deepens and the periods become longer. Stars in the middle of the strip have the highest amplitudes, periods longer than 300\,s, and many combination frequencies resulting from nonlinear interactions among the modes caused by changes in the convection zone during the pulsation \citep{1992MNRAS.259..519B}. \citet{2005ApJ...633.1142M} showed how these nonlinearities can be used to measure the depth of the convection zone and constrain mode identification, a critical precursor for asteroseismic inference. These stars in the middle of the instability strip also have modes that show amplitude and frequency variability, unlike the hotter ZZ\,Cet stars which have stable modes. Finally, as the stars approach the red edge of the DAV instability strip around $11\,000$\,K, increased radiative damping overwhelms the convective driving   \citep{2018ApJ...863...82L}.

As the DAV white dwarfs cool even further towards the red edge of the instability strip, some of them undergo outbursts, a remarkable new phenomenon first discovered by \citet{2015ApJ...809...14B} in a 1.5-yr {\it Kepler} light curve of KIC\,452982; \citet{2017ASPC..509..303B} listed six DAVs that outburst. They found that the outbursts release $10^{26} - 10^{27}$\,J causing an increase in brightness of the star by about 15\%, recur irregularly on a timescale of days, and cause the pulsation mode amplitudes and frequencies to vary. \citet{2018ApJ...863...82L} suggested that the outbursts are connected with resonant three-mode coupling among overstable parent modes and (otherwise) damped daughter modes. This is potentially testable with studies of combination terms in the frequencies of outbursting DAV stars. This outburst phenomenon is now also seen in as-yet unpublished TESS data, including a star that is also an eclipsing binary with an orbital period of 1.5\,hr. It will be interesting to see whether binarity has an importance in the outbursting DAV phenomenon, and whether the stars have any relation to cataclysmic variables. Large numbers of outbursting DAV stars are expected to be discovered with the ZTF and VRO surveys, to be followed up from the ground and with short-cadence TESS, and later PLATO, observations. Since the outbursts last for $4 - 24$\,hr, they may have been present in ground-based studies, but were lost in the data processing which has to correct for daily artifacts. Some reprocessing of ground-based multisite campaigns may resolve this issue, but it is the long time spans of the space data that will illuminate this phenomenon.

\citet{2020A&A...638A..82B} reported on TESS first year observations for 18 previously known ZZ\,Cet stars, comparing the TESS results with ground-based results, with some increase in the number of detected modes for 8 of the stars. The other 10 did not show detectable pulsations from a combination of reduced amplitude because of the 2-min cadence, field crowding because of the relative large TESS pixels (21\,arcsec), and low count rates for the faster stars. This provides some caution for future use of TESS for faint objects, although the more recent 20-s cadence is a great benefit for white dwarf studies. 

\subsection{GW\,Vir stars (DOV)}

The GW\,Vir stars are the hottest class of pulsating white dwarfs and pre-white dwarfs. They are DO white dwarfs, classified so by the presence of HeII lines in their spectra; they are H deficient, and rich in C and O. They are driven by the $\kappa$-mechanism where the variable opacity comes from partial ionization of C and O in the upper layers of the stars. The prototype GW\,Vir, also known as PG\,1159,  is one of the best-studied of all pulsating white dwarfs. \citet{1991ApJ...378..326W} found over 100 mode frequencies, determining the mass to high precision, the rotation period, and limiting the magnetic field strength. \citet{2008A&A...477..627C} extended that to nearly 200 mode frequencies, improving the mass determination and magnetic field limit, and measuring the extent of the mode trapping region. Both of these studies used data from the Whole Earth Telescope (WET). No GW\,Vir stars were observed by the {\it Kepler} mission, but now 6 known GW\,Vir and PNNV stars have been observed by TESS  \citep{2021A&A...645A.117C}. That work includes PNNV that are also Wolf-Rayet (WC) stars, which have similar pulsation characteristics to the GW\,Vir stars. These stars will become even hotter as they contract, reaching $T_{\rm eff} > 200\,000$\,K before evolving down the white dwarf cooling tracks. The study of \citet{2021A&A...645A.117C} confirmed, with more mode frequencies, earlier ground-based studies of some stars, determined the rotation period in one star, and showed that the mass determinations from period spacings are robust, and that the asteroseismic distances agree well with Gaia distances. These hottest pre-white dwarfs are in transition from being the cores of red giants to spending the rest of time (unless the proton has a finite lifetime) on the white dwarf cooling tracks, eventually becoming black dwarfs.  

\begin{marginnote}[]
\entry{GW\,Vir stars include} \ DOV white dwarfs; g\,mode pulsators. $T_{\rm eff}$ of $75,000 - 100\,000$\,K;  periods of $300- 2600$\,s. 

\vspace{2mm}
Planetary Nebulae Nuclei Variables (PNNV); g\,mode pulsators. $T_{\rm eff}$ of $100\,000 - 200\,000$\,K;  periods of $400- 6000$\,s.  
\end{marginnote}

\subsection{V777\,Her stars (DBV)}
\label{dbv}

The DBV stars show He\,I lines and little or no H in their spectra. Winget (1982) predicted the existence of this class theoretically, then \citet{1982ApJ...262L..11W} successfully discovered the prototype, GD 358 = V777 Her. \citet{2017ApJ...835..277H} reported a study of rotation in the hottest ($T > 30\,000$\,K) known DBV star, PG\,0112+104 using a 79-d Kepler K2 data set. This star shows a clear, stable 10.17-hr rotational light curve from a spot (or spots) that is reminiscent of the typical rotational light curves of the magnetic spotted $\alpha^2$\,CVn Ap stars. This was the first known spotted, pulsating white dwarf star. There are 26 detected pulsation frequencies from 11 independent dipole and quadrupole modes, which can be used to probe radial differential rotation, since the dipole and quadrupole modes have different pulsation cavities; that study is yet to be done. The low amplitudes suggest that more such stars will be found in TESS data, with surface spots, and dipole and quadrupole modes that will provide better understanding of the radial dependence of rotation in white dwarfs, as has been done in many main-sequence stars (see \citealt{2019ARA&A..57...35A}). 

\begin{marginnote}[]
\entry{V777\,Her stars} \ DBV white dwarfs; g\,mode pulsators. $T_{\rm eff}$ of $22\,000 - 32\,000$\,K;  periods of $120- 1080$\,s. 
\end{marginnote}

\citet{2019A&A...632A..42B} have reported on a first study of a DBV star with 2-min cadence TESS data showing the benefit of the long data sets for stars with stable frequencies. The best asteroseismic temperatures from their models were somewhat discordant with the spectroscopic temperature; further modeling should resolve that. With 20-s TESS we can expect significant new constraints  on white dwarf structure.

\subsection{Pulsating Extremely Low Mass Variable (ELMV) white dwarf stars}
\label{elmwd}

Low mass (LM) white dwarfs are now known in relatively large numbers as a consequence of targeted surveys. They are generated when mass transfer in binary evolution exposes the core of a red giant before the He flash. There is a subset these stars at Extremely Low Mass (ELM) with $M < 0.18$\,M$_\odot$. \citet{2012ApJ...750L..28H} discovered the first pulsating ELM (ELMV), SDSS\,J184037.78+642312.3, a DAV star with a mass of only $0.17$\,M$_\odot$.  Adiabatic pulsation analysis showed that these stars pulsate in g\,modes that primarily probe the core region, unlike higher mass DAV stars where the g\,modes have pulsation cavities largely confined to the non-degenerate atmosphere. \citet{2013A&A...557A..19A} calculated evolutionary models and determined mass and cooling age for 58 ELM stars, three of which pulsate in g\,modes.

\begin{marginnote}[]
\entry{Low mass (LM) white dwarfs} \ He core white dwarfs with $M \simkl 0.43$\,M$_\odot$; remnant red giant cores stripped prior to the He flash. 
\entry{Extremely Low mass variables (ELMV)} \ He core, H atmosphere white dwarfs with $M < 0.18$\,M$_\odot$ with surface temperatures in the range $7800 - 10\,000$\,K; g\,mode pulsators with periods of $100- 6300$\,s.
\end{marginnote}

\citet{2016A&A...585A...1C} performed a non-adiabatic pulsation analysis of binary evolutionary sequences for ELMV stars, starting with a 1.0-M$_\odot$ main-sequence star with a 1.4-M$_\odot$ neutron star companion. They found large numbers of excited p\,modes and g\,modes driven by the $\kappa - \gamma$ mechanism, i.e. excitation by time variable opacity enhanced by radial depth dependence of the adiabatic exponent in the ionization zone. They also found some short period g\,modes to be unstable via the $\epsilon$-mechanism providing driving in the H shell burning zone, as there is sufficient residual H in ELM white dwarfs for some nuclear energy generation. That these ELM white dwarfs can still sustain some H fusion via the p-p chain lengthens their cooling time by orders of magnitude compared to slightly more massive white dwarfs with $M > 0.18 - 0.20$\,M$_\odot$ where multiple CNO flashes burn off the residual H quickly. \citet{2017A&A...607A..33C} performed asteroseismic analyses for 9 ELMV stars using those models. Each star has only a few mode frequencies, whereas the models suggest many modes should be excited. Forward modeling gave masses consistent with spectroscopic masses. 

\subsection{Blue Large Amplitude Pulsators (BLAPs)} 

BLAPs are a newly discovered class of pulsators found in both the OGLE-IV and ZTF surveys. These stars are stripped red giant cores that pulsate with periods in the range $180-2400$\,s, with amplitudes of a few percent. While there are only 18 currently known, they seem to split into two groups of lower ($4.2 \le \log g \le 4.6$) and higher ($5.3 \le \log g \le 5.7)$ surface gravity. \citet{2020MNRAS.492..232B} and  \citet{2021MNRAS.tmp.1916B} have modeled their evolutionary stage, concluding that they are pre-ELM H-shell burning white dwarfs. The pulsations are driven by the Fe-peak opacity enhanced by the radiative levitation of Fe for stars in both gravity ranges, so they are likely the same phenomenon.  Although some BLAPs overlap with the sdBV pulsators (Section\,4.2) in the HR~Diagram, they have larger amplitudes than the sdBVs, hence are not confused with them. There is discussion about the nomenclature for this class, which could therefore change. They are beautiful example of a new class of a late evolutionary stellar evolution stage where asteroseismology will inform our understanding of red giant cores. It is expected that many more of these stars will be found in large surveys, such as the ZTF and VRO. 

\begin{marginnote}[]
\entry{Blue Large Amplitude Pulsators (BLAPs)} \ pre-ELM H-shell burning white dwarfs; g\,mode pulsators with periods of $180- 2400$\,s.
\end{marginnote}

\subsection{DQV, hot DAV, and Ultra-massive DAV stars}

After the main classes, DOV, DBV, and DAV, there are three additional classes of variable white dwarfs. The DQ white dwarfs, which have H and He deficient, C rich atmospheres; the hot-DAV white dwarfs with temperatures at the hot end of the DBV instability strip; and the ultra-massive (UM) DAV white dwarfs which have excellent potential to study crystallization in detail \citep{2019A&A...621A.100D}. While pulsation remains a possible explanation for the DQV stars, evidence is strong that their variability is rotational, arising from a spot or spots in the presence of a strong magnetic field \citep{2016ApJ...817...27W}. In that case, no asteroseismic inference is possible for these stars, but their rapid rotation (periods in the range minutes to 10s of minutes) carries information about their origin and precursors. Those are thought to be either AGB single stars or mergers of double-degenerate binary stars. In the AGB scenario the spread in rotation rates for DQV stars must be a function of internal angular momentum transfer in the precursor during its evolution, and magnetic braking once the white dwarf is formed. TESS is targeting these stars to build up a broader pictures of their variability. 

 \begin{marginnote}[]
\entry{DQV white dwarfs} \ White dwarfs with H and He deficient, C rich atmospheres. It is uncertain whether their variability is pulsational or rotational.
\entry{Hot DAV white dwarfs} \ H atmosphere white dwarfs at the hot edge of the DB instability strip.
\entry{UM DAV white dwarfs} \ Ultra-Massive DAV white dwarfs with masses in the  range $1.10 - 1.29$\,M$_\odot$. Strongly crystallized C-O or O-Ne cores.
\end{marginnote}

The hot DAV stars are H atmosphere stars at the hot edge of the DB (helium atmosphere) instability strip. \citet{2005EAS....17..143S} predicted their existence, following which three cases were found (\citealt{2013MNRAS.432.1632K}; \citealt{2008MNRAS.389.1771K}). Some uncertainty whether these stars constitute a new class of pulsating white dwarfs persists, but the recent discovery of a fourth such star, HE\,1017--1352, by \citet{2020MNRAS.497L..24R} strengthens the case. For the latter star there are TESS data for which the pulsation signal, while not at the 1/1000 level of False Alarm probability, is consistent with the robust detections from ground-based photometry. The lower significance of the TESS signal is a result of the lack of blue light in the TESS passband, which demonstrates why ground-based photometric observations of hot pulsating white dwarfs are still needed in the space photometry era. 

The adiabatic pulsation properties of the UM DAV ($1.10 \le M \le 1.29$\,M$_\odot$) white dwarfs were studied in depth theoretically by \citet{2019A&A...621A.100D}, who found the cores to be highly crystallized and suggested that g\,mode period spacings may be able to discern stars with C-O cores from those with O-Ne cores, an intriguing prospect. 

\subsection{EL\,CVn binary systems}
\label{elcvn}

\citet{2013Natur.498..463M} discovered a new kind of pulsator in a ELM white dwarf star in an eclipsing binary, 1SWASP\,J024743.37-251549.2. This pulsator is a pre-He white dwarf, the core of a stripped red giant star, in orbit with an A or F star. It pulsates in mixed modes, which have g\,mode character in the core and p\,mode character in the outer layers, hence they probe the entire star. \citet{2014MNRAS.437.1681M} found 17 of these systems and named the class after the brightest of those, EL\,CVn. \citet{2015ApJ...803...82R} reported on two of the {\it Kepler} systems (out of 13 found in {\it Kepler} data) with mass and radius determinations, finding thermal bloating of the pre-white dwarf to the size of a typical red dwarf, up to 14 times what the cold degenerate radius would be.  \citet{2018MNRAS.475.2560V} found 36 EL\,CVn systems in a machine learning search of 800\,000 stars with the ZTF, and more than 100 of these systems are now known from a variety of surveys. They found the pre-He white dwarfs to have temperatures of $7900 - 17\,000$\,K, while their AF companions have $6600 - 10\,000$\,K. They also found a secure mass range of $0.12 - 0.28$\,M$_\odot$, but possibly as high as $0.5$\,M$_\odot$, and radii of $0.17 - 0.65$\,R$_\odot$ -- very bloated for white dwarfs. 
 
\begin{marginnote}[]
\entry{EL\,CVn binaries} \ Bloated pre-He white dwarfs in short-period ($< 3$\,d) orbits about an A or F star that in some cases is also a $\delta$~Sct star.
\end{marginnote}

 \citet{2020MNRAS.499L.121L} showed that EL\,CVn stars are almost all in hierarchical triples with the orbital period of the eclipsing pair less than 3\,d. The pulsation signal, if any, from the LM or ELM white dwarf is strongly diluted by the light of the AF primary, which is in many cases a $\delta$~Sct star. \citet{2020ApJ...888...49W} analyzed two EL\,CVn systems with TESS data finding pulsations in both the $\delta$~Sct primary and the pre-He white dwarf. The frequency ranges of these two types of variables are mostly separated in frequency space, so both can independently be studied asteroseismically. 

The EL\,CVn stars are important for understanding how close binaries with white dwarfs form. This is a field that has developed in the last 10 years from ground-based wide-field survey data, primarily SuperWASP\footnote{https://wasp-planets.net}, and now is benefiting from long duration light curves from space data. At present only two EL\,CVn systems have pulsating white dwarf components, but with the expansion of discovery these systems, particularly by ZTF and VRO, asteroseismic study will complement inference from orbital analyses.

\section{Extreme Horizontal Branch stars: The hot subdwarf O and B variables}
\label{app:ehbstars}

In the early 1950s Jesse Greenstein -- the founder of the astronomy program at CalTech -- was observing white dwarfs, sdB, and sdO stars with the coud\'e spectrograph on the then-largest optical telescope, the Palomar 200-inch (5-m). At IAU Symposium 3 held in Dublin in September 1955 he commented, ``As far as I know, light variations have not been found in the white dwarfs'' \citep{1957IAUS....3...41G}. He finished his conference report by noting, ``Observations designed to detect possible short-period light variations [in white dwarfs and subdwarfs] would be desirable.'' Greenstein recognized then that the lack of detection of photometric variations could well be that the amplitudes were just too small for the precision then obtainable. He was correct. A decade later in the 1960s variations began to be found in white dwarfs and subdwarfs. \citet{1968ApJ...153..151L} discovered pulsation for the first time in a white dwarf, HL\,Tau\,76. For the subdwarf stars the discovery of the class of p\,mode sdBV pulsators now known as V361\,Hyi (or EC\,14026) stars only came in the late 1990s, more than 40 years after Greenstein's suggestion. 

In a case of theory leading observational discovery (as for the DBV white dwarfs; Section\,\ref{dbv}), \citet{1996ApJ...471L.103C} predicted sdB stars  to pulsate from models. They were then independently discovered the following year in the Edinburgh-Cape Blue Object Survey (\citealt{1997MNRAS.285..640K}, \citealt{1997MNRAS.285..645K}, \citealt{1997MNRAS.285..651S}, \citealt{1997MNRAS.285..657O}). Just a few years later \citet{2003ApJ...583L..31G} serendipitously discovered a new class of sdBV g\,mode pulsators now known as V1093\,Her stars. For unknown reasons the incidence of pulsation for the V361\,Hyi stars is only about 10\%, while that for the V1093\,Her is about 75\%. Both classes have pulsation driven by the $\kappa$-mechanism operating on Fe and Ni opacity \citep{1996ApJ...471L.103C}  where abundances of those elements have been enhanced in the driving zone by radiative levitation (\citealt{2001PASP..113..775C}, \citealt{2006MNRAS.372L..48J}). \citet{2014A&A...569A.123B} performed detailed calculations of both radiative levitation and gravitational settling for 9 elements showing that Fe and Ni are enhanced by factors up to 60 and 4000, respectively, in the driving zone, naturally accounting for both the p\,mode and g\,mode pulsations and the extents of the instability strips.

\begin{marginnote}[]
\entry{V361\,Hyi stars} \ subdwarf B variables, sdBV stars, pulsating in p\,modes with periods of $60 - 800$\,s. sdBV$_{\rm r}$ stars \\ (r for rapid). $28\,000 \le T_{\rm eff} \le 45\,000$\,K.
\entry{V1093\,Her stars} \ sdBV stars pulsating in g\,modes with periods of $0.5 - 4$\,hr. sdBV$_{\rm s}$ stars \\ (s for slow). $23\,000 \le T_{\rm eff} \le 30\,000$\,K.
\entry{DW\,Lyn stars} \ hybrid sdBV stars pulsating in both p\,modes and  g\,modes. \\ sdBV$_{\rm rs}$ stars.
\entry{V366\,Aqr stars} \ sdBV stars with enhanced Sr, Y, Zr pulsating in g\,modes. $T_{\rm eff} \sim 36\,000$\,K. 
\end{marginnote}

Asteroseismology of sdBV stars has provided the fundamental parameters mass and radius, as well as structural parameters -- the envelope mass, the core mass and the core composition -- illuminating their evolutionary origins. Their masses are concentrated near 0.47\,M$_\odot$, supporting the stripped red giant origin for these stars.  Asteroseismology of the much rarer sdOV stars is still developing \citep{2019arXiv190500654V}. \citet{2017MNRAS.467.3963K} and \citet{2019MNRAS.485.4330K} discovered an sdOV star, EC\,03089-6421, with $T_{\rm eff} = 52\,000$\,K and $M \sim 0.1$\,M$_\odot$, pulsating with very short periods of 27\,s, 31\,s, and 34\,s. The pulsations in this star are potentially driven by the $\epsilon$-mechanism and it may be a progenitor of an ELM white dwarf. Finding pulsations driven by the $\epsilon$-mechanism is a long-standing goal, with previous claims or suggestions that have not been confirmed. It will be exciting if models can confirm $\epsilon$-mechanism excitation in EC\,03089-6421, giving a direct asteroseismic constraint on the nuclear energy generation in this star.

There is an intriguing class of heavy metal sdBV stars, the V366\,Aqr stars, with only three members known. They have temperatures in the range of the V361\,Hyi p\,mode pulsators, yet they pulsate in g\,modes. Strikingly, they show overabundances of Ge, Sr, Y, and Zr, the latter up to $10^4$ solar. The standard Fe-Ni opacity driving mechanism does not explain how sdB stars this hot pulsate in g\,modes, rather than p\,modes. The $\epsilon$-mechanism has been proposed, but ruled out for the third member, PHL\,417, by \citet{2020MNRAS.499.3738O} based on the stability of the highest amplitude mode. Hence, the driving mechanism for the V366\,Aqr stars is not yet known. 

\citet{2021FrASS...8...19L} reviewed asteroseismology of hot subdwarf stars, discussing individually many of the sdBV stars observed by the main {\it Kepler} and K2 missions. The TESS mission is now observing many more of these stars, and studies are in progress. \citet{2021arXiv210515137U} introduced about 40 sdBV stars observed by TESS in 2-min and 20-s cadence and analyzed 5 V1093\,Her stars pulsating exclusively in g\,modes. No rotational multiplets were found among primary dipole modes, and the main asteroseismic inference was that standard convective core mixing provides good agreement with the mean period spacings. 

That is not in agreement with \citet{2019A&A...632A..90C} who analyzed a single TESS sdBV star with detailed asteroseismic inference. They found a He core mass larger than in standard models in this star and three others previously analyzed. They concluded that further understanding of the physical processes that govern the core structure are needed. They also found a precise mass of $0.39 \pm 0.01$\,M$_\odot$, which is lower than the mean sdB mass of $0.47$\,M$_\odot$, suggesting that the star evolved from a red giant that had not ignited He in a core flash. They derived radius, luminosity and  a double-layered He and H envelope structure as a consequence of gravitational settling. Their best models have a core with $0.20 \pm 0.01$\,M$_\odot$ with 43\% of the He already burned. They found a rough estimate of the core O mass fraction, which ultimately may constrain the $^{12}{\rm C}(\alpha,\gamma)^{16}{\rm O}$ reaction rate. These results illustrate the power of asteroseismology for sdBV stars. 

\citet{2007Natur.449..189S} used an $O-C$ analysis of pulsation timing ({Section\,\ref{app:fm-pm}) for the sdBV star V391\,Peg, which shows both p\,modes and g\,modes. They found a cyclic variation in the pulsation frequencies and inferred the presence of a planet with a minimum mass of 3.2\,M$_{\rm Jup}$ in a 3.2-yr orbit. They inferred that the planet would have been about 1\,AU from the progenitor star when it was on the main-sequence, suggesting that planets orbiting at that distance may survive the red giant stage, with implications for the fate of Earth. Later, \citet{2018A&A...611A..85S} found with more extensive data that the presence of the planet is uncertain, since the oscillatory variations in two of the frequencies may not be consistent. 

The rewards from detecting planets orbiting pulsating stars are great, but this star is an additional warning about the interpretation of pulsational frequency variability (Evolutionary weather; Sections\,2.2). That is also true for attempts to measure evolutionary frequency changes. \citet{2016A&A...594A..46Z} give an explanation for these shorter time-scale frequency variations in the sdB star KIC\,10139564. They find amplitude and frequency modulations that are the clear signature of three-mode non-linear resonant coupling. \citet{2016A&A...585A..22Z} also find similar nonlinear mode couplings in the DBV white dwarf KIC\,8626021, for which they emphasize the complications for trying to measure white dwarf cooling times. 

The number of known subdwarfs is increasing dramatically, as with other classes of asteroseismic targets. \citet{2020A&A...635A.193G} has provided a catalog of nearly 6000  hot subdwarfs. There is also a contentious possible new class of subdwarfs, the sdA stars. \citet{2016MNRAS.455.3413K} recently classified a group of stars as sdA based only on their hydrogen dominated spectra and surface gravities, $5.5 \le \log g \le 6.5$, showing they lie below the main-sequence. \citeauthor{2016MNRAS.455.3413K} proposed that many of them are ELM white dwarfs (Section\,\ref{elmwd}), but \citet{2017ASPC..509..453H} strongly disputed that, and their status is yet to be clarified.

\section{New techniques: FM, PM, and sNa}

\subsection{The frequency modulation (FM) and phase modulation (PM) techniques}
\label{app:fm-pm}

It has long been recognized that stable pulsation frequencies can be used as frequency standards to study orbital motion in pulsating stars. Traditionally, this was exploited in $O-C$ diagrams, where, typically, the observed ($O$) time of pulsation maximum of a star minus a chosen computed ($C$) value was plotted against time. A straight line in this plot means the period is constant over the duration of observations; any other shape indicates variations in the period. For example, for steady evolutionary period changes a parabola is expected in the diagram, whereas for orbital motion with a circular orbit a sinusoid is expected. Information about the orbit can be extracted, just as is done using spectroscopy to obtain orbital radial velocities. 

In examining large numbers of FTs for stars in the {\it Kepler} data with 4\,yr of observations, I recognized that we were seeing stars where all pulsation frequencies were appearing as equally spaced multiplets. That could happen asteroseismically if all modes were nonradial with a full complement of azimuthal $m$ modes excited. But the patterns did not look like that. The theoretical explanation was presented by \citet{2012MNRAS.422..738S}, with a first application to the multiply-eclipsing quintuplet system HD\,181469 (KIC\,4150611), which has a $\delta$~Sct/$\gamma$~Dor star as the largest mass component \citep{2017A&A...602A..30H}. We called the method FM, for frequency modulation, and further development \citep{2015MNRAS.450.3999S} showed how radial velocities can be obtained without spectra, which are exceedingly expensive in terms of telescope time. This means binary star orbits can now be completely solved with only photometric time series for pulsating stars with stable pulsation frequencies. Such stars are common and the method is being exploited, particularly for $\delta$~Sct stars.

However, many $\delta$~Sct stars show non-periodic amplitude modulation \citep{2016MNRAS.460.1970B}, and that broadens and distorts the peaks in the FTs which interferes with measuring the frequency multiplets produced by FM. A solution to that problem was developed by \citet{2014MNRAS.441.2515M,2016MNRAS.461.4215M}. It is a version of the $O-C$ diagram. As binary stars orbit their barycentre, the light travel time from the star to Earth varies, causing the pulsation phase to vary, thus they called the technique PM, for phase modulation. That can be used to extract the variable time delays and a plot of those against time in a new kind of $O-C$ diagram. They showed that all of the information about a binary star orbit that is available from spectroscopic radial velocities is available from PM. Stars with one pulsating component are, therefore, called PM1 (in analogy to SB1, single-lined spectroscopic binaries) and stars with two pulsations components are similarly PM2. \citet{2018MNRAS.474.4322M} then examined over 2000 main-sequence A and F stars, discovering 341 non-eclipsing binary stars, 24 of which were PM2, tripling the number of stars on this region of the main-sequence with orbital solutions. 

The PM method is an application of pulsating stars to the related field of study of the orbits of binary stars, hence it provides fundamental data, such as masses, rather than asteroseismic inference on the interiors of those stars. The method is not affected by amplitude modulation of the pulsations, since that does not affect the pulsation phase; in this way it is superior to the FM method. It does require binning the data into time chunks -- \citet{2018MNRAS.474.4322M} used 10-d sections of the light curves for this -- which puts a lower limit of 20\,d on the orbital period determined. FM uses the entire data set, hence has better frequency resolution, and can thus explore periods shorter than 20\,d. PM is also automated, whereas FM needs hands-on reductions. With an additional 10 binary systems \citet{2020MNRAS.493.5382M} used the PM method to explore the orbits of 351 binary stars with orbital periods up to 1500\,d in the {\it Kepler} data -- largely unexplored parameter space since, for geometric reasons, such long periods are unusual in eclipsing binaries. The 1500-d limit, is, of course, a consequence of the duration of the main {\it Kepler} mission. Future, longer duration photometric missions will lengthen the orbital periods that can be studied, filling in the gap between eclipsing binary and visual binary studies. 

The FM and PM methods are complementary. They obviate the need for telescope time to obtain spectra over full orbits, hence save vast amounts of money and astronomers' time. \citet{2018MNRAS.474.4322M} estimated that their study of 341 binaries saved about 100\,yr of medium to large telescope time! The FM and PM methods provide binary star fundamental parameters to complement, and to check, asteroseismic determinations in the same stars. And they show asteroseismologists the patterns expected in FTs from orbital motions, so that those frequency multiplets are not mistaken for rotationally split frequency multiplets. 

\subsection{Super-Nyquist Asteroseismology (sNa)}
\label{app:sna}

In examining the frequency multiplets in {\it Kepler} data that led to the FM and PM techniques, I noticed that some pulsating stars show evidence of frequency triplets indicative of orbital motion, but only for some frequencies, not all. \citet{2013MNRAS.430.2986M} showed that this is the result of the periodic modulation of the observing time by the light travel time effect of the {\it Kepler} satellite orbit around the Sun. The {\it Kepler} mission had cadences that were equally spaced in time on board, thus  not equally spaced in barycentric time. That lifts the Nyquist ambiguity in the FT, greatly extending the number of stars that can be studied asteroseismically in {\it Kepler} long-cadence data. An example application is the discovery by \citet{2019MNRAS.488...18H} of 6 new roAp stars with correctly identified pulsation frequencies in the super-Nyquist frequency range in the {\it Kepler} data. \citet{2018arXiv181110205S} have proposed how the technique may be applied to data from the future PLATO mission. 

\bibliographystyle{ar-style2.bst}
\bibliography{araa_kurtz}

\end{document}